%% file: main.tex
\documentclass{article}
\input defi

\renewcommand{\margin}[1]{}

\newcommand{\siball}[1]{}

\newcommand{\sibsmall}[1]{#1}

\title{A Proof Procedure for Hybrid Logic with Binders, Transitivity
  and Relation Hierarchies\\
(extended version)}

\author{Marta Cialdea Mayer}
\date{Universit\`a di Roma Tre, Italy} 
\begin{document}
\maketitle

\begin{abstract}
In  previous works, a tableau calculus has been defined, which
constitutes a decision procedure for hybrid logic with the converse
and global modalities and a restricted use of the binder. This work
shows how to extend such a calculus to multi-modal logic enriched with
features largely used in description logics:
transitivity and relation inclusion assertions.

The separate addition of  either transitive relations or
relation hierarchies to the considered decidable fragment of
multi-modal hybrid logic can easily be shown to stay decidable, by
resorting to results already proved in the literature. However, such
results do not directly allow for concluding whether the logic
including both features is still
decidable.
The existence of a terminating, sound and complete calculus for the
considered logic proves that the addition of transitive relations and
relation hierarchies to such an expressive decidable fragment of hybrid
logic does not endanger decidability.

A further result proved in this work is that the logic extending the
considered fragment with the addition of 
graded modalities (the modal counterpart of number restrictions of
description logics)  has an
undecidable satisfiability problem, unless further syntactical
restrictions are placed on the universal graded modality. 

\end{abstract}

\input intro

\input logic
\input undec

\input preprocessing

\input calcolo

\input examples
\input nobinder

\input concl
\bibliography{hybiblio}
\appendix
\input appendix

\input termina
\input completeness

\end{document}

%% file: defi.tex
\usepackage{amsthm}
\usepackage{epsfig,proof,latexsym,amsmath}
\usepackage{enumitem,url,amsfonts,amssymb,scalefnt}
\usepackage{stmaryrd}
\usepackage{epsfig}
\usepackage{epstopdf}

\newcommand{\butta}[1]{}
\newcommand{\margin}[1]{\marginpar{\tiny #1}}

\newtheorem{theorem}{Theorem}
\newtheorem{lemma}{Lemma}
\newtheorem{definition}{Definition}
\newtheorem{example}{Example}

\newtheorem{newlemma}{Lemma}

\newtheorem{defi}{Definition}

\newcounter{prop}
\newenvironment{properties}
        {\begin{list}%
          {P\arabic{prop}.}%
          {\usecounter{prop}
           \setlength{\rightmargin}{\leftmargin}}
        }
        {\end{list}}   

\newcommand{\R}{R}
\renewcommand{\S}{S}
\renewcommand{\P}{T}
\renewcommand{\r}{r}
\newcommand{\s}{s}
\newcommand{\p}{t}

\newcommand{\allora}{~\Longmapsto}
\newcommand{\inv}{\textsf{Inv}}
\newcommand{\clo}{\operatorname{Clo}}
\newcommand{\produce}{\rhd}
\newcommand{\B}{\mbox{\bf B}}

\newcommand{\Subf}{\operatorname{Subf}}
\newcommand{\Cmp}{\operatorname{Cmp}}
\newcommand{\binder}{\mbox{$\mathord\downarrow$}}
\newcommand{\leadstobinder}{\leadsto^{\downarrow}}
\newcommand{\at}{\mbox{$\mathord :\,$}}
\newcommand{\M}{\mbox{$\cal M$}}
\newcommand{\N}{\mbox{$\cal N$}}
\newcommand{\F}{\mbox{$\cal F$}}
\newcommand{\ie}{\mbox{i.e.}}
\newcommand{\Ie}{\mbox{I.e.}}
\newcommand{\wrt}{\mbox{w.r.t.}}

\newcommand{\freccione}{\prec}
\def\Nat{{\rm I\!N}}
\newcommand{\Hl}{\textsf{HL}}
\newcommand{\A}{\textsf{A}}
\newcommand{\E}{\textsf{E}}
\newcommand{\leadstoA}{\leadsto^{\scalefont{0.7}{\textsf{A}}}}

\newcommand{\indietro}{\hspace{-5ex}}
\newcommand{\PROP}{\mathsf{PROP}}
\newcommand{\NOM}{\mathsf{NOM}}
\newcommand{\X}{\mathsf{VAR}}
\newcommand{\ramo}{\mbox{$\cal B$}}
\newcommand{\REL}{\mbox{$\mathsf{REL}$}}
\newcommand{\subrole}{\mbox{$\,\sqsubseteq\,$}}
\newcommand{\trans}{\mathsf{Trans}}
\newcommand{\calA}{\mbox{$\cal A$}}
\newcommand{\link}{\mathsf{Link}}

\newcommand{\rel}{\mathsf{Rel}}
\newcommand{\comma}{,\allowbreak}
\newcommand{\framcalcolo}{\Hl_m(@\comma \binder\comma \E\comma 
\Diamond^-\comma \trans\comma \subrole)
  \setminus\binder\Box}
\newcommand{\frammento}{\Hl(@\comma \binder\comma \E\comma \Diamond^-)
  \setminus\Box\binder\Box}
\newcommand{\miniframmento}{\Hl(@\comma \binder\comma \E\comma \Diamond^-)
  \setminus\binder\Box}
\newcommand{\mmframmento}{\Hl_m(@\comma \binder\comma \E\comma \Diamond^-)
  \setminus\Box\binder\Box}
\newcommand{\thframmento}{\Hl_m(@\comma \binder\comma \E\comma 
\Diamond^-\comma \trans\comma \subrole)
  \setminus\Box\binder\Box}
\newcommand{\fhl}{\Hl(@\comma \binder\comma \E\comma \Diamond^-)}
\newcommand{\mmfhl}{\Hl_m(@\comma \binder\comma \E\comma \Diamond^-)}
\newcommand{\thfhl}{\Hl_m(@\comma \binder\comma \E\comma \Diamond^-
\comma \trans\comma \subrole)}

\newcommand{\sibyl}{\textsf{Sibyl}}
\newcommand{\imp}{\rightarrow}

%% file: intro.tex
\section{Introduction}

Hybrid languages are extensions of modal logic that allow for naming
and accessing states of a structure explicitly (see, for instance,
\cite{areces-tencate}). Their main
distinguishing feature is represented by special atomic propositions,
called {\em nominals}, which give names to states: a nominal is true
in exactly one state of the model.  The two operators specific of
hybrid languages are the {\em satisfaction operator} ($@$), allowing
for jumping to a point named by a nominal, regardless of the
accessibilities in the structure, and the {\em  binder}
($\binder$), allowing for dynamically binding {\em state variables} to states
and  referring to these states later on.

Other modal operators can be added to the basic hybrid language, such
as the converse modalities ($\Diamond^-$ and
$\Box^-$) and the global ones ($\E$ and $\A$).  Moreover,  hybrid
languages can rely on a 
multi-modal base, allowing for modelling structures
with different accessibility relations. In this case, the basic
modalities $\Diamond$ and $\Box$ (and their converses,  if present) are
indexed by relation symbols.  
Hybrid multi-modal languages can also be enriched with
a feature largely used in description logics, {\ie} 
the possibility of declaring an accessibility relation to be
transitive  and/or included in another one.   

In this work, basic hybrid logic (with nominals
only, beyond the modal operators $\Diamond$ and $\Box$) will be
denoted by $\Hl$, and basic multi-modal hybrid logic by
$\Hl_m$. Logics extending $\Hl$ or $\Hl_m$ with operators
$O_1,\dots,O_n$ (and their duals) are denoted by $\Hl(O_1,\dots,O_n)$
and $\Hl_m(O_1,\dots,O_n)$, respectively.  Multi-modal languages
including transitivity assertions and/or relation hierarchies are
denoted in the same way, just including $\trans$ (for transitivity)
and/or $\subrole$ (for relation inclusion) among $O_1,\dots,O_n$.

The satisfiability problem for formulae of any hybrid logic
$\Hl(O_1,\dots,O_n)$ or $\Hl_m(O_1,\allowbreak \dots,\allowbreak O_n)$, 
where $O_i\in\{@,\Diamond^-,\E\}$, is decidable
\cite{areces-tencate}.  Unfortunately, due to the high expressive
power of the binder, $\Hl(\binder)$ is undecidable
\cite{roadmap,blackburn-seligman}.

There are both semantic and syntactic restrictions allowing for
regaining decidability of hybrid logic with the binder.  Restricting
the frame class is a way of restoring decidability.
For instance, 
 decidability of $\Hl(@,\binder)$
\margin{modif}
can be recovered by restricting the frame
class to frames of bounded width ({\ie} frames where the number of
successors of each state is bounded by a natural number) \cite{FranBis}.
However,  the interplay
with multi-modalities (or the addition of other operators) is not
always harmless.  For instance, $\Hl(\binder)$ over transitive frames
is decidable \cite{MSSW10}, but $\Hl_m(\binder)$ and $\Hl(@,\binder)$ 
are not \cite{MSSW10,MS07hylo06}.

\butta{
\margin{modif} "For instance, decidability of .... can be
recovered by restricting the frame class to frames of bounded width
(i.e. frames where the number of successors of each state is bounded)
[25]." I'm not able to find such a claim in [25]. In fact, Theorem 4
of [25] shows the opposite of what you seem to claim: The
satisfiability problem of $H(@,\binder)$ on $K_\omega$ 
is recursively enumerable but
not decidable.
}

In \cite{FranBis} it is proved that the satisfiability problem for
formulae in $\Hl(@\comma \binder\comma \E\comma \Diamond^-)$ is
decidable, provided that their negation normal form contains no
universal operator ({\ie} either $\Box$ or $\Box^-$ or $\A$) scoping
over a binder, that in turn has scope over a universal operator.  Such
a fragment of hybrid logic is denoted by $\frammento$. The result is
proved by showing that there exists a satisfiability preserving
translation of $\frammento$ into $\miniframmento$, {\ie} the set of
formulae in negation normal form where no universal operator occurs in
the scope of a binder.  The standard translation of hybrid logic into
first order classical logic \cite{roadmap,FranBis} maps, in turn,
formulae in $\miniframmento$ into universally guarded formulae, that
have a decidable satisfiability problem \cite{Gradel98onthe}.

Decidability of {$\mmframmento$} can be proved by the same reasoning,
and the separate addition of either relation hierarchies or transitive
relations can easily be shown to stay decidable.
Since also the translation of a relation inclusion axiom is a guarded
formula, the reduction argument used in
\cite{FranBis} shows that $\Hl_m(@,\binder,\E,\Diamond^-,\subrole)
  \setminus\Box\binder\Box$ has a decidable satisfiability problem.
But transitivity axioms make the guarded fragment (GF) of first order
logic undecidable  \cite{Gradel98onthe}. 
On the other side, if transitive
relations only occur in guards, GF is decidable
\cite{GF+transitivity}, and consequently so is
 satisfiability in $\Hl_m(@\comma \binder\comma
\E\comma \Diamond^-\comma \trans)  
  \setminus\Box\binder\Box$.
But the translation of relation inclusion axioms may have transitive
relations outside guards. Therefore, 
in the presence of both transitive
relations and relation hierarchies, the decidability question cannot
be settled by resorting to results already
proved in the literature.

\butta{
 by reduction to the
first order guarded fragment and by resorting to results already
proved in the literature \cite{GF+transitivity}.  However, such
results do not directly allow for concluding whether the logic
including both features is still decidable.
}

The above reported 
arguments showing decidability of fragments of hybrid logic with
binders are all of semantical nature. The first proof procedures
constituting satisfiability decision procedures for such fragments are
defined in \cite{frammento-tab2011,jar2012}.
  In
particular, \cite{jar2012} presents a tableau based 
satisfiability decision procedure for $\frammento$, and such a procedure
is  extended to multi-modal hybrid logic {$\thframmento$} in
\cite{cade2013}.
This work is a revised and extended version of \cite{cade2013},
including full proofs and a new result concerning the graded modalities.
 A
tableau calculus is presented, which terminates and is sound and
complete for formulae in the fragment
$\Hl_m(@,\binder,\E,\Diamond^-,\trans,\subrole) \setminus\binder
\Box$, {\ie} formulae in negation normal form where no 
universal operator occurs in the scope of a binder, with the addition of
transitivity assertions and relation hierarchies.  A preprocessing
step along the lines of \cite{FranBis}  turns the calculus into a
satisfiability decision procedure for the fragment $\thframmento$.
Soundness, completeness and termination of the tableaux calculus thus
imply that the satisfiability problem for the fragment of multi-modal
hybrid logic $\thframmento$ is decidable.  

The language of {$\thframmento$} 
allows for representing some interesting frame properties.
For instance, if 
 transitivity assertions have the form $\trans(\r)$, where $\r$ is a
relation symbol,  and  inclusion
assertions have either the form $\s\subrole \r$ ($\s$ is a sub-relation
of $\r$) or $\s^-\subrole \r$ (the inverse of $\s$ is a sub-relation
of $\r$), the following frame properties can be represented:
\butta{
 (in this
work we use the notation $t\at F$ for satisfaction formulae, meaning
that $F$ holds in the state denoted by the nominal or variable $t$):
}
\medskip

\hspace*{-1em}
\begin{tabular}{ll}
Transitivity: & 
 $\trans(\r)$\\
Symmetry:  
&$\r^-\subrole \r$ \\
Reflexivity:  
~~& $\A\binder x.\Diamond_{\r} x$\\
At most $n$ states: &  $\E\binder x_1.\dots \E\binder x_n.
\A(x_1\vee \dots\vee x_n) $\\
At least one 
$\r$-sibling: & $\A\binder x. \Diamond_{\r}^-\Diamond_{\r}\neg x$\\
At least $n$ 
$\r$-successors:
&$\A\binder x.\Diamond_{\r} \binder y_1.(x:\Diamond_{\r}(\neg y_1\wedge
                        \binder y_2.$\\
&~~~~~~~~~~~~~~~~~($x:\Diamond_{\r}(\neg y_1\wedge \neg
                        y_2\wedge \binder y_3.\dots))))$
\end{tabular}\medskip\\
\butta{
(Trichotomy means $\forall x \forall y (R(x,y)\vee x=y\vee R(y,x))$,
and universality means $\forall x \forall y R(x,y)$).
Invece \cite{b00-internalizing} definisce ``at most 2
states'' con una formula pura che (usando invece le globali) dice:
$\A\binder x.\A\binder y.\A(\neg x\wedge\neg y \rightarrow  x:y)$
perch\'e deve usare formule senza $\E$ (l'uso di nominali in un
assioma puro \`e come avere un $\A\binder$ che li ``lega'').

Number restrictions generali:

 Therefore satisfiability in 
$\mmframmento$ is decidable over 
frames enjoying any combination of the above properties.
}

Restricted uses of the binder are of interest also also in the context
of description
logics \cite{DescriBinder,stepmother}.
Considering that $\Hl_m(@\comma \binder\comma \E\comma 
\Diamond^-\comma \trans\comma \subrole)$  subsumes the description
logic $\cal SHOI$,
a natural question arises: can number restrictions (or, in modal
terms, graded modalities) be added to {$\thframmento$} without
endangering 
decidability? In this work we show that the answer is, in general,
 negative: the
satisfiability problem for 
hybrid logic with either the satisfaction operator or 
the converse modalities, functional
restrictions and binders, without the critical pattern
$\Box\binder\Box$, is undecidable. However, decidability can be preserved 
by placing additional syntactical 
\margin{modif}
restrictions on the occurrences of the graded modalities.

The work is organized as follows. Section \ref{logic} defines the
syntax and semantics of $\Hl_m(@\comma \binder\comma \E\comma 
\Diamond^-\comma \trans\comma \subrole)$. 
The undecidability result for the extension of the considered fragment
of hybrid logic by means of the graded modalities is proved in Section
\ref{undec}, that also shows how to further restrict the occurrences
of such modalities so as to preserve decidability. 
The satisfiability
preserving translation of formulae in $\mmframmento$ 
into $\Hl_m(@\comma \binder\comma \E\comma \Diamond^-)
  \setminus\binder\Box$ is presented in Section
\ref{preprocess}.  Section \ref{calcolo} is the core of this work,
describing the terminating tableau calculus for formulae 
in $\Hl_m(@\comma \binder\comma \E\comma \Diamond^-\comma\trans\comma \subrole)
  \setminus\binder\Box$. The termination and completeness proofs are
  given in the Appendix. Section \ref{examples} illustrates the
  calculus in action by means of some examples and 
Section \ref{nobinder} briefly compares the binder free subsystem of the
calculus with other works.\siball{
Section \ref{sibyl} contains a sketchy description of
the  prover called
    {\sibyl},
an implementation
of the proof procedure, and }
Section \ref{concl} concludes this work.

%% file: logic.tex
\section{Syntax and semantics of multi-modal hybrid logic with
  transitive relations and inclusion assertions}
\label{logic}

  Well-formed expressions of $\thfhl$ are
partitioned into two categories: {\em formulae}  
(for which the metasymbols $F,G,H$ 
are used) and {\em assertions}.
%
\butta{
{\em Formulae} are built out of a set $\PROP$ of propositional
letters, a set $\NOM$ of nominals, an infinite set $\X$ of state
variables, and a set $\REL$ of relation symbols (all such sets being
mutually disjoint), and defined by the following grammar:
\[\begin{array}{ll}
F:= & p ~\mid~ a  ~\mid~ x ~\mid~\neg F ~\mid~ F \wedge F  ~\mid~ F \vee F
 ~\mid~ \Diamond_R F~\mid~ \Box_R F\\
& \mid~ \Diamond^-_{R} F~\mid~ \Box^-_{R} F ~\mid~\E F~\mid~\A F
~\mid~ a\at F ~\mid~ x\at F ~\mid~\binder x. F
\end{array}\]
where $p \in \PROP$, $a \in \NOM$, $x \in \X$ and $R\in\REL$.  In this
work, the notation $t\at F$ is used (for $t\in\NOM\cup\X$) rather than
$@_t F$.}
The language is based on a set $\PROP$ of propositional
letters, a set $\NOM$ of nominals, an infinite set $\X$ of state
variables, and a set $\REL$ of relation symbols (all such sets being
mutually disjoint).
\margin{modif} 
When using a meta-symbol $\r$ for a relation symbol in $\REL$, 
the corresponding uppercase letter, $\R$, will denote a {\em relation}, {\ie}
either the relation denoted by
$\r$ itself (a {\em forward relation}) or its converse, denoted by $\r^-$ (a
{\em backward relation}). A backward relation  $\r^-$
is  the set of pairs of states $\langle w,w'\rangle$ such that
$\langle w',w\rangle$ is in the relation denoted by $\r$.
\butta{
If $\r\in\REL$ is a relation symbol, the
corresponding uppercase letter, $\R$, denotes a {\em relation}, {\ie}
either $\r$ itself (a {\em forward relation}) or its converse $\r^-$ (a
{\em backward relation}).
A backward relation  $\r^-$
is intended to denote the set of pairs of states $\langle w,w'\rangle$ such that
$\langle w',w\rangle$ is in the relation denoted by $\r$.}
\butta{
\margin{Modif} If r ∈ REL is a relation symbol, the corresponding
uppercase letter, R, denotes a relation,....  This is a bit
confusing. Since r is a *symbol*, it can not both be true that "R
denotes a relation" and that R can be "r itself". I think it is
incorrect to say that "R denotes a relation".
}

{\em Formulae} are
defined by the following grammar:
\[
F:=  p \,\mid\, u   \,\mid\,\neg F \,\mid\, F \wedge F  \,\mid\, F \vee F
 \,\mid\, \Diamond_\R F\,\mid\, \Box_\R F
 \mid\,\E F\,\mid\,\A F
\,\mid\, u\at F  \,\mid\,\binder x. F
\]
where $p \in \PROP$, $u \in \NOM\cup\X$, $x \in \X$ and $\R$ is either
a forward or
backward relation.  In this
work, the notation $u\at F$ is used  rather than
$@_u F$.
  The metavariables $a,b,c,d$ are used for
nominals, $x,y,z$ for state variables and $\r,\s,\p$ for relation symbols
(every metavariable possibly decorated by subscripts and quotes).

If $F$ is a formula, $x$ a state variable and $a$ a nominal, then
$F[a/x]$ denotes the formula obtained from $F$ by substituting $a$ for
every free occurrence of $x$ (an occurrence of $x$ is free if it is
not in the scope of a $\binder x$).  If $a_0,\dots,a_n,b_0,\dots,b_n$
are nominals, then $F[b_0/a_0,\dots,b_n/a_n]$ denotes the formula
obtained from $F$ by simultaneously replacing $b_i$ for every
occurrence of $a_i$.

{\em Assertions} are either {\em transitivity assertions,} of the form
$\trans(\r)$, for $\r\in\REL$, or {\em inclusion assertions,} of either
form $\r\subrole\s$ or $\r^-\subrole \s$, for $\r,\s\in\REL$.  Note that
backward relations are allowed only on the left of the $\subrole$
symbol.  This is only a syntactical restriction, and expressions of
the form $\R\subrole \S$ are used as abbreviations of their
semantically equivalent assertions: $\r^-\subrole \s^-$ stands for 
$\r\subrole \s$, and $\r\subrole \s^-$ for $\r^-\subrole \s$.
\butta{
Similarly, $\trans(\R)$ abbreviates $\trans(\r)$ in both cases
$\R=\r$ and $\R=\r^-$.
since $R^-\subrole
S^-$ is equivalent to $R\subrole S$, and $R\subrole S^-$ is equivalent
to $R^-\subrole S$.
}

An {\em interpretation} $\cal M$ of an $\thfhl$ language is a tuple
$\langle W, \rho, N, I\rangle$ where $W$ is a non-empty set (whose
elements are the {\em states} of the interpretation), $\rho$ is a
function mapping every $\r\in\REL$ to a binary relation on $W$
($\rho(\r)\subseteq W \times W$), $N $ is a function $\NOM\rightarrow
W$ and $I$ a function $W \rightarrow 2^{\PROP}$.
The following abbreviation will be used:
\[
w \R w' =
\left\{
\begin{array}{ll}
\langle w,w'\rangle\in\rho(\r) & 
\mbox{if }\R=\r\mbox{ is a forward
  relation}\\
\langle w',w\rangle\in
\rho(\r)& \mbox{if }\R=\r^-\mbox{ is a backward  relation}
\end{array}\right.
\]
\butta{
write $w \R w'$ as a shorthand for $\langle w,w'\rangle\in
\rho(\r)$ if $\R=\r$, and $w R^- w'$ for $\langle w',w\rangle\in
\rho(R)$.
}

If ${\cal M}=\langle W, \rho, N, I \rangle$ is an interpretation,
$w\in W$, $\sigma$ is a variable assignment for ${\cal M}$ ({\ie} a
function $\X\rightarrow W$) and $F$ is a formula, the relation
$\M_w, \sigma \models F$ is inductively defined as follows:

\begin{enumerate}
\item  $\M_w, \sigma \models p$ if $p\in I(w)$, for $p\in \PROP$.
\item  $\M_w, \sigma \models a$ if $N(a)=w$, for $a\in\NOM$.
\item  $\M_w, \sigma \models x$ if $\sigma(x)=w$, for $x \in \X$.
\item $\M_w, \sigma \models \neg F$ if $\M_w, \sigma \not \models F$.
\item $\M_w, \sigma \models F\wedge G$ if $\M_w, \sigma \models F$ and 
$\M_w, \sigma \models G$.
\item $\M_w, \sigma \models F \vee G$ if either $\M_w, \sigma \models
  F$ or $\M_w, \sigma \models G$. 

\item $\M_w, \sigma \models a\at  F$ if $\M_{N(a)}, \sigma \models F$, for 
$a \in \NOM$. 
\item $\M_w, \sigma \models x\at  F$ if $\M_{\sigma(x)}, \sigma \models F$,
for $x \in \X$.
\item  $\M_w, \sigma \models \binder x. F$ if
 $\M_{w}, \sigma^w_x \models F$, where $\sigma^w_x$ is 
 the variable assignment such that $\sigma^w_x(x)=w$ and, for $y\neq
 x$, 
$\sigma^w_x(y)=\sigma(y)$.
\item $\M_w, \sigma \models \Box_\R F$ if for every $w'$ such that
  $w\R w'$,
$\M_{w'}, \sigma \models F$.
\item $\M_w, \sigma \models \Diamond_\R F$ if there exists  $w'$ such
that  $w\R w'$ 
and $\M_{w'}, \sigma \models F$.

\item $\M_w, \sigma \models \A F$ if $\M_{w'}, \sigma \models F$ 
for all $w'\in W$.
\item $\M_w, \sigma \models \E F$ if 
$\M_{w'},\sigma \models F$ 
for some $w'\in W$.
\end{enumerate}

A formula $F$ is {\em satisfiable} if there exist an interpretation
$\cal M$, a variable assignment $\sigma$ for ${\cal M}$ and a state
$w$ of $\M$, such that $\M_w, \sigma \models F$.
Two formulae $F$ and $G$ are logically equivalent 
when, for every
interpretation $\cal M$, assignment $\sigma$ and state $w$ of $\M$:
$\M_w, \sigma \models F$ if and only if $\M_w, \sigma \models G$.
 A
formula $F$ holds in a state $w$ of a model $\M$ ($\M_w\models F$) iff
$\M_w, \sigma \models F$ for every variable assignment $\sigma$.

Every formula in $\mmfhl$ is logically equivalent to a formula in
negation normal form (NNF), where negation appears only in front of
atoms.
Therefore, considering only formulae in NNF does not restrict the
expressive power of the language.

If $\calA$ is a set of assertions, an interpretation $\langle
W, \rho, N, I  \rangle$ is a
model of $\calA$ if:
\begin{enumerate}
\item for all  $\r\in\REL$ such that $\trans(\r)\in\calA$, 
 $\rho(\r)$ is a transitive relation;

\item for all $\r,\s\in\REL$, if 
$\r\subrole \s\in\calA$, then $\rho(\r)\subseteq \rho(\s)$;

\item for all $\r,\s\in\REL$ and all $w,w'\in W$, if
$\r^-\subrole \s\in\calA$ and 
$\langle w,w'\rangle\in \rho(\r)$, then $\langle w',w\rangle\in \rho(\s)$. 
\end{enumerate}

Finally, if $F$ is a formula and $\calA$ a set of assertions,
$\{F\}\cup\calA$ is satisfiable if there exist a model $\M$ of $\calA$
and a state $w$ of $\M$ such that $\M_w\models F$.

%% file: undec.tex
\butta{
\newcommand{\dx}{R} 
\newcommand{\up}{U} 
\newcommand{\grid}{G} 
}
\newcommand{\dx}{\mathsf{r}}
\newcommand{\up}{\mathsf{u}}
\newcommand{\grid}{\mathsf{g}}
\newcommand{\Dx}{\mathsf{R}}
\newcommand{\Up}{\mathsf{U}}
\newcommand{\Grid}{\mathsf{G}}
\newcommand{\esse}{\mathsf{S}}
\newcommand{\exactly}{\textsf{Exactly}}

\section{The graded modalities}
\label{undec}

The logic introduced in Section \ref{logic}
 subsumes, in modal terms, the
description logic $\cal SHOI$.  The latter does not include number
restrictions, one of the important expressive constructs of
description logics.  Therefore, a natural question arises: is it
possible to add the modal counterpart of number restrictions ({\ie}
 {\em graded modalities})  to the
fragment $\mmframmento$ without endangering decidability?  This
section 
shows that, although a restricted use of graded modalities can be
added to the fragment,  the general answer to this question is
negative.

The  graded
  modalities are here denoted by  
$\Diamond_{\r}^n$ and  $\Box_{\r}^n$, where $n\in\Nat$.
 In the presence of the converse
 modalities, also graded modalities indexed by backward relations can
 be allowed, so their general forms 
are $\Diamond_{\R}^n$ and  $\Box_{\R}^n$.
Their semantics  is the following:
\begin{itemize}
\item   $\M_w, \sigma \models \Diamond_{\R}^n F$ iff there are at least
  $n+1$ distinct states $w_1,...,w_n$ such that $wRw_i$ and
$\M_{w_i},\sigma\models F$. 
\item  $\M_w, \sigma \models \Box_{\R}^n F$ iff there are at most 
  $n$ distinct states $w_1,...,w_n$ such that $wRw_i$ and
$\M_{w_i},\sigma\not\models F$.

\end{itemize}

When considering the interplay between the binder and universal
modalities in order to tackle decidability issues, 
 the
universal graded modality $\Box_{\R}^n$ is to be included, with $\Box_{\R}$ and
$\A$, among the universal modalities (with the obvious consequence on
the meaning of the patterns $\Box\binder\Box$ and $\binder\Box$).
The first part of this section shows
 how to restrict the use of the graded modalities 
\margin{modif}
so as to
obtain a decidable sublogic of
$\Hl_m(@\comma \binder\comma \E\comma \Diamond^-\comma \Diamond^n)
\setminus \Box\binder\Box$.   
The second part proves that, in general, the satisfiability problem
for
$\Hl_m(@\comma \binder\comma \E\comma \Diamond^-\comma \Diamond^n)
\setminus \Box\binder\Box$ 
is undecidable.

\butta{
\margin{modif}
      The expressive power of ... is actually the same as
..., since the graded modalities can be expressed in terms of the
binder." Please refer back to p. 3 where you show how to do this, or
refer forward 
to the proof of Thm 1.
}
The expressive power of
$\Hl_m(@\comma \binder\comma \E\comma \Diamond^-\comma \Diamond^n)$ is
actually the same as  
$\Hl_m(@\comma \binder\comma \E\comma \Diamond^-)$, since
the graded modalities can be expressed in terms of the binder (see
the proof of Theorem \ref{graded-dec} below). 
The limitations on 
the use of 
 $\Box_{\R}^n$, in order to keep a decidable satisfiability problem,
 are however stronger than those required for the other
universal modalities. 
\margin{modif}
Moreover, occurrences of the existential graded
modality have to be restricted, too.

\butta{
\margin{modif}
Vincoli su $\Diamond_R^n$: 
either $\Diamond_R^n F$ does not occur in the scope of a universal
modality, or $F$ does not contain any universal modality.
}
\margin{modif}
\begin{theorem}
\label{graded-dec}
The satisfiability
problem for a formula $G$ (in NNF) belonging to the fragment
$\Hl_m(@\comma \binder\comma \E\comma \Diamond^-\comma \Diamond^n)
\setminus
\Box\binder\Box$ is decidable provided
that:
\begin{enumerate}
\item for every subformula $\Box_{\R}^nF$ of $G$:
\begin{enumerate}
\item $\Box_{\R}^nF$ does not occur in the scope of any universal
modality;
\item  $F$ does not contain the pattern $\binder\Box$.
\end{enumerate}
\item for every subformula $\Diamond_{\R}^nF$ of $G$, 
either $\Diamond_R^n F$ does not occur in the scope of a universal
modality, or $F$ does not contain any universal modality.
\end{enumerate}
\end{theorem}

\begin{proof}
The proof shows how to express the graded modalities as abbreviations
of formulae which,  under the additional restrictions 1 and 2, do not
contain the pattern $\Box\binder\Box$.

The
existential graded modality can easily be expressed as an abbreviation
of a formula in $\Hl_m(@\comma \binder\comma \E\comma \Diamond^-)$:
$\Diamond_R^n F$ is  
equivalent to the formula $(\Diamond_{\R}^nF)^*$
defined below, where the state variables $x,y_1,\dots,y_n$ do not
occur free in $F$.
\[
\begin{array}{lcl}
(\Diamond_{\R}^0 F)^* & \equiv_{def} & \Diamond_{\R} F\\
(\Diamond_{\R}^1 F)^* & \equiv_{def} & \binder x.\Diamond_{\R} (F\wedge \binder
y_1. x\at \Diamond_{\R}(F\wedge \neg y_1)             \\
(\Diamond_{\R}^2 F)^* & \equiv_{def} & \binder x.\Diamond_{\R} (F\wedge \binder
y_1. x\at \Diamond_{\R}(F\wedge \neg y_1\wedge \binder y_2.
x\at \Diamond_{\R}(F\wedge \neg y_1\wedge \neg y_2))             \\
\dots\\
(\Diamond_{\R}^n F)^* & \equiv_{def} \\
\multicolumn{3}{l}
{\begin{array}{l}
 ~~~\binder x.\Diamond_{\R} (F\wedge\\
~~~~~~~~~~~~ \binder
y_1. x\at \Diamond_{\R}(F\wedge \neg y_1\wedge\\
~~~~~~~~~~~~~~~~~~~~~~~~~~ \binder y_2.x\at \Diamond_{\R}(\dots \wedge\\
~~~~~~~~~~~~~~~~~~~~~~~~~~~~~~~~~~~~~~~~\binder y_{n-1}.
x\at \Diamond_{\R}(F\wedge \neg y_1\wedge\dots\wedge \neg y_{n-1}\wedge\\
~~~~~~~~~~~~~~~~~~~~~~~~~~~~~~~~~~~~~~~~~~~~~~~~~~~~~~
 \binder y_n.
x\at \Diamond_{\R}(F\wedge \neg y_1\wedge \dots\wedge \neg y_n))\dots)))
\end{array}}
\end{array}
\]
It is easy to see that, if $F$ belongs to 
$\mmframmento$, then so does 
 the formula
 $(\Diamond_{\R}^n F)^*$.
\margin{modif} If however $(\Diamond_{\R}^n F)^*$ occurs as a
subformula of a formula $G$, in order for $G$ to belong to the
considered fragment, either no universal operator must scope over
$(\Diamond_{\R}^n F)^*$, or $F$ must contain no universal operators.

Considering that
 $\Box_{\R}^n F\equiv \neg \Diamond_{\R}^n \neg F$, 
the universal graded modality can obviously be expressed in terms of
the binder, too. However, 
the NNF of  $\neg (\Diamond_{\R}^n \neg F)^*$ contains
the critical pattern $\Box\binder\Box$, so that resorting to the
definition of $\Box_{\R}^n$ in terms of $\Diamond_{\R}^n$ is of no
help to the aim of establishing decidability results for the hybrid language
including the graded modalities.

However, $\Box_{\R}^n F$ can also be defined in a different way (here
again, it is assumed that   the  variables $x,y_1,\dots,y_n$ do not
occur free in $F$):
\[
\begin{array}{lcl}
(\Box_{\R}^0F)^* &\equiv_{def} & \Box_{\R} F\\
(\Box_{\R}^1F)^* &\equiv_{def} & \Box_{\R} F\vee\binder x.\Diamond_{\R}(\binder y_1.
          x\at\Box_{\R}(F\vee y_1))\\
(\Box _{\R}^2F)^* &\equiv_{def} & \Box_{\R} F\vee\binder x.\Diamond_{\R}(\binder y_1.
          x\at\Diamond_{\R}(\binder y_2.
          x\at\Box_{\R}(F\vee y_1\vee y_2)))\\
\dots\\
(\Box_{\R}^nF)^*  & \equiv_{def} &\Box_{\R} F\vee\\
&&~\binder x.\Diamond_{\R} (\binder
y_1. x\at \Diamond_{\R}(\binder y_2.x\at\Diamond_{\R}(\dots\\
&&~~~~~~~~~~~~~~~~~~~~~~~~~~~~~~~~~~
 \binder y_n.
x\at \Box_{\R}(F\vee y_1\vee \dots\vee y_n))\dots)))
\end{array}
\]

\noindent
The following reasoning  shows that $(\Box _{\R}^nF)^*$ is equivalent to $\Box
_{\R}^nF$.
\begin{enumerate}
\item Let us assume that
$\M_w,\sigma\models (\Box _{\R}^nF)^*$. Then one of the following cases holds:
\begin{enumerate}
\item  $\M_w,\sigma\models\Box_{\R} F$; then there are no states $w'$
  such that $w\R w'$ and $\M_{w'},\sigma\not\models F$, so that
  trivially $\M_w,\sigma\models\Box _{\R}^nF$.
\item $\M_w,\sigma^w_x\models\Diamond_{\R} (\binder
y_1. x\at \Diamond_{\R}(\binder y_2.x\at\Diamond_{\R}(\dots
 \binder y_n.
x\at \Box_{\R}(F\vee y_1\vee \dots\vee y_n))\dots)))$: there exist
(not necessarily distinct) states $w_1,\dots,w_n$ such that 
$w\R w_i$ and 
$\M_w,\sigma^{w,w_1,\dots,w_n}_{x,\,y_1,\,\dots,\,y_n}\models
\Box_{\R}(F\vee y_1\vee\dots\vee y_n)$.  Consequently, for every state
$w'$ such that $w{\R}w'$ and 
$\M_{w'},\sigma^{w,w_1,\dots,w_n}_{x,\,y_1,\,\dots,\,y_n}\not\models F$,
$w'\in\{w_1,\dots,w_n\}$.
Since the variables $x,y_1,\dots,y_n$ do not occur free in $F$, this
amounts to saying that 
for every state
$w'$ such that $w{\R}w'$ and 
$\M_{w'},\sigma\not\models F$,
$w'\in\{w_1,\dots,w_n\}$:
 there are at most $n$ distinct states $w_1,\dots,w_n$
such that $w{\R}w_i$ and $\M_{w_i},\sigma\not\models F$.
\end{enumerate}
\item For the converse, let us assume that
$\M_w,\sigma\models \Box _{\R}^nF$, and that there are exactly $k\leq
  n$ distinct states $w_1,\dots,w_k$
such that $w{\R}w_i$ and $\M_{w_i},\sigma\not\models F$.
Let us consider the following cases:
\begin{enumerate}
\item $k=0$. Then $\M_w,\sigma\models\Box_{\R} F$, hence 
$\M_w,\sigma\models (\Box _{\R}^nF)^*$.

\item $k>0$. Let then $n=k+m$, for $m\geq 0$, and
$w_1,\dots,w_k,w_{k+1},\dots,w_{k+m}$ be the sequence of $n$ states where
$w_1,\dots,w_k$ are followed by $m$ repetitions of $w_k$. Since $k>0$,
such a sequence is well defined.

For all $i=1\dots n$, $w{\R}w_i$, $\M_{w_i},\sigma\not\models F$, and for every
state $w'$ such that $w{\R}w'$ and $\M_{w'},\sigma\not\models F$,
$w'=w_i$ for some $i=1\dots n$.
Since there are no free occurrences of $x,y_1,\dots,y_n$ in $F$, 
$\M_{w_i},\sigma\not\models F$ is equivalent to
$\M_{w_i},\sigma^{w,w_1,\dots,w_n}_{x,\,y_1,\,\dots,\,y_n}\not\models F$.
As a consequence, 
$\M_{w_i},\sigma^{w,w_1,\dots,w_n}_{x,\,y_1,\,\dots,\,y_n}\models
x\at\Box_{\R}(F\vee y_1\vee\dots\vee y_n)$.

Since moreover, for all $i=1\dots n$, $w{\R}w_i$, 
$\M_{w_i},\sigma^{w,w_1,\dots,w_n}_{x,\,y_1,\,\dots,\,y_n}\models
\Diamond_{\R} (\binder
y_1.\allowbreak  x\at \Diamond_{\R}(\binder y_2.x\at\Diamond_{\R}(\dots
 \binder y_n.\allowbreak x\at \Box_{\R}(F\vee y_1\vee \dots\vee y_n))\dots)))$. 
Since $y_1,\dots,y_n$ do not occur free in $F$,
$\M_{w_i},\sigma^{w}_{x}\models
x\at\Diamond_{\R} (\binder
y_1. x\at \Diamond_{\R}(\binder y_2.x\at\Diamond_{\R}(\dots
 \binder y_n.\allowbreak x\at \Box_{\R}(F\vee y_1\vee \dots\vee
 y_n))\dots)))$, so  
$\M_{w_i},\sigma\models (\Box_{\R}^n)^*$.

\end{enumerate}
\end{enumerate}

In order for $(\Box_{\R}^nF)^*$ to belong to the
considered decidable fragment of $\Hl$, 
its subformula $F$ must not
contain the pattern $\binder\Box$ (which would occur in the scope of
$\Box_{\R}$). Moreover, $\Box_{\R}^nF$ itself must not occur in
the scope of a universal modality.
The statement of the theorem is 
 a direct consequence of the above considerations.
\end{proof}

If no further restrictions are placed on the graded modalities, the
satisfiability problem for
$\Hl_m(@\comma \binder\comma \E\comma \Diamond^-\comma \Diamond^n)\setminus 
\Box\binder\Box$ is undecidable.
In order to establish this fact, only functional restrictions are
required.

Obviously,
  $\Box_{\r}^1$  can be used to  express
 functionality of the relation $\rho(r)$:
 $\M_w, \sigma \models \Box_{\r}^1 \bot$ iff 
$w$ has at most one
  $\r$-successor.
Analogously,    $\Box_{\r^-}^1$ can be used to express injectivity:
 $\M_w, \sigma \models \Box_{\r^-}^1 \bot$ iff 
$w$ has at most one
  $\r$-predecessor.
Formulae of the form $\Box_\R^1\bot$ will be called {\em functional
restrictions}, and, if $\R$ is a forward relation, they are called
 {\em forward functional
restrictions}, otherwise {\em
   backward functional restrictions}.

In what follows,
$\Hl_m(@\comma \binder\comma\Diamond^1)$
denotes the hybrid multi-modal language with the satisfaction
operator, the binder and forward
functional restrictions.
And
$\Hl_m(@\comma\binder\comma\Diamond^1)\setminus \Box\binder \Box$
denotes the fragment of 
$\Hl_m(@\comma \binder\comma\Diamond^1)$ consisting of formulae 
whose NNF
do 
not contain any occurrence of the binder
that is both in the scope and has in its scope a universal modality
({i.e.} either $\Box_{\r}$ or 
$\Box_{\r}^1$).

\begin{theorem}\label{undec-at}
The satisfiability problem for 
$\Hl_m(@,\binder,\Diamond^1)\setminus \Box\binder
  \Box$
is undecidable.
\end{theorem}

\begin{proof}
The proof is  based on
a modification of the encoding of the $\Nat\times\Nat$ tiling problem
presented 
in \cite{FranBis} in order to prove that $\Hl$ with binders is
undecidable, even in the absence of the satisfaction operator.
That proof  is conceived so as to highlight that
 the source of undecidability
is the presence of the pattern $\Box\binder\Box$. 
Here, we show how the formulae containing the critical pattern can be
replaced by use of number restrictions and the satisfaction operator.

Let us first briefly recall what  the 
 $\Nat\times\Nat$ tiling problem is.
A tile is a square with fixed orientation and each edge coloured from
a finite set of colours. A set of tile types tiles a space if tiles of
the given types
can cover the space, in such a way that adjacent tiles have the same
colour on the matching sides. The
$\Nat\times\Nat$ tiling problem is then: given a finite set of tile
types $T$, can the infinite grid $\Nat\times\Nat$ be tiled using only
tiles of the types in $T$? This problem is well known to be undecidable
(see, e.g., \cite{tiling}).

The $\Nat\times\Nat$ tiling problem can be reduced to
 the satisfiability problem for
$\Hl_m(@,\binder,\Diamond^1)\setminus \Box\binder
  \Box$ with  three modalities: $\Diamond_{\up}$ (to move one step
up in the grid), $\Diamond_{\dx}$ (to move one step to the right in the
grid), and $\Diamond_{\grid}$ (to reach all the points of the grid),
interpreted by the accessibility relations $\Up$, $\Dx$ and $\Grid$,
respectively. Let $T$ be a finite set of tiles, and for each tile $t\in
T$ let $left(t)$, $right(t)$, $top(t)$, and $bottom(t)$ denote the four colors
of $t$. We will now give a hybrid formula $\pi_T$ that describes a
tiling of $\Nat\times\Nat$ using the tile types in $T$,
and does not contain
the pattern $\Box\binder\Box$. The formula $\pi_T$ is the conjunction
of the following formulae:

\begin{description}
\item[Spypoint.] $\alpha$ is the conjunction of the following
  formulae:, where $a$ is a nominal:
\[\begin{array}{l}
 a \wedge \Diamond_{\grid} a\wedge
 \Box_{\grid}\Diamond_{\grid}   a \\
 \Box_{\grid}\Box_{\up}\binder x.(\Diamond_{\grid}(a\wedge \Diamond_{\grid} x))\\
\Box_{\grid}\Box_{\dx}\binder x.(\Diamond_{\grid}(a\wedge \Diamond_{\grid} x))
\end{array}
\]  
\item[Functionality.] 
$\beta = 
\Box_{\grid}\Diamond_{\up}\top \wedge 
\Box_{\grid}\Diamond_{\dx}\top \wedge 
\Box_{\grid}\Box_{\up}^1 \bot \wedge 
\Box_{\grid}\Box_{\dx}^1 \bot
$.

\item[Grid.] 
$\gamma = \Box_{\grid}\binder x.\Diamond_{\up}\Diamond_{\dx}\binder y.
x:\Diamond_{\dx}\Diamond_{\up} y$.\\

\item[Tiling] $\delta=\Box_{\grid}(\delta_1\wedge\delta_2\wedge \delta_3)$, 
where
\[\begin{array}{l}
\delta_1 = \bigvee_{t\in T}(p_t \wedge \bigwedge_{t'\in T:t\neq
  t'}\neg p_{t'})\\
\delta_2 = \bigwedge_{t\in T}(p_t\imp \Box_{\dx}\bigvee_{t'\in
  T:left(t')=right(t)} p_{t'})\\
\delta_3 = \bigwedge_{t\in T}(p_t\imp \Box_{\up}\bigvee_{t'\in
  T:bottom(t')=top(t)} p_{t'})
\end{array}\]
 \end{description}

The formula $\alpha$ 
exploits the ability of the binder to force the existence of a
``spypoint'' (the state denoted by $a$), from which
 the entire grid can be accessed via the relation $\Grid$.
It says that the current state is
  named $a$, that is $\Grid$-related to itself.
  Each $\Grid$-successor of $a$ has $a$ as a $\Grid$-successor (every
  point in the grid sees $a$ via $\Grid$). And
the set of $a$'s $\Grid$-successors (the points in the grid) 
is closed under $\Up$ and $\Dx$.
This formula is the same as the corresponding one in \cite{FranBis}.

$\beta$  says that 
all states in the submodel induced by the spypoint $a$ (all points of
the grid) have exactly  an $\Up$
successor and an $\Dx$ successor, {i.e.} $\Dx$ and $\Up$ are total functions.
It is worth pointing out that  the universal graded modality occurs in the scope
of a universal modality in the formula $\beta$.
The formula used in \cite{FranBis} to express functionality of $\Up$ contains
instead the pattern $\Box\binder\Box$:
\[\Box_{\grid}\Diamond_{\up}\top \wedge \Box_{\grid}\binder x.\Box_{\grid}(s\imp
\Box_{\grid}(\Diamond_{\up} x\imp \Box_{\up} x)\]
(and analogously for the relation $\Dx$).

$\gamma$  states that for
 every state $x$ of the grid, it is possible to go up ($\Up$) and
then right ($\Dx$) to a state $y$ that can also be accessed from $x$ by
moving first right  and then up.
The same ``grid property'' is expressed in \cite{FranBis} by a formula that does
not contain the satisfaction operator, but contains
 the critical pattern:
\[\Box_{\grid}\binder x.\Box_{\grid}(s \imp \Box_{\grid}(\Box_{\up}\Box_{\dx}\neg x
  \vee \Box_{\dx}\Box_{\up} x))\]

$\delta$ states that the
    grid is well-tiled: formula $\delta_1$ states
    that exactly one tile is placed at each node of the grid, $\delta_2$ says
    that horizontally adjacent tiles must match, and $\delta_3$ says that
    vertically adjacent tiles must match.
$\delta$ is the same formula  as the corresponding one in
    \cite{FranBis}.

In order to reduce the $\Nat\times\Nat$ tiling problem to the
satisfiability of $\pi_T=\alpha\wedge\beta\wedge\gamma\wedge\delta$,
it must be shown that, for any set of tile types $T=\{t_1,\dots,t_n\}$, $T$
tiles $\Nat\times\Nat$ iff
    the  formula $\pi_T$
 is satisfiable.  The proof is quite standard, and is outlined below.

Suppose that $\M_{w_0}\models \pi_T$, 
for $\M=\langle W, \rho, N, I \rangle$, 
and let $w\esse w'$ abbreviate
$\langle w,w'\rangle \in \esse$ for   $\esse \subset W\times W$.
 We show how to define a function
$tile:\Nat\times\Nat\imp T$ that is a tiling of $\Nat\times\Nat$.

Let $Grid=\{w\in W\mid  w_0\Grid w\}$ 
be the set of grid points.
Since $\M_{w_0}\models a\wedge \Diamond_{\grid} a\wedge 
\Box_{\grid}\Diamond_{\grid} a$ 
($\alpha$), $N(a)=w_0$, 
$Grid\neq \emptyset$, and for every $w\in Grid$, 
$w\Grid w_0$.

Since moreover $\M_{w_0}\models \Box_{\grid}\Box_{\up}\binder x.
(\Diamond_{\grid}(a\wedge
\Diamond_{\grid} x))$, every point $w$ that is an $\Up$-successor of
some $w'\in Grid$ is such that, if $\sigma(x)=w$, then $\M_w\sigma\models 
\Diamond_{\grid}(a\wedge \Diamond_{\grid} x)$, thus
$w_0\Grid w$, {\ie} 
 $w\in
Grid$. And the same holds for $\Dx$: every $\Dx$-successor of a point in
the grid is in the grid.

$\M_{w_0}\models \beta$ 
implies that every $w\in Grid$ has at least one $\Up$-successor and an
$\Dx$-successor, and for every $w,w_1,w_2\in Grid$, if
\butta{
$\langle w,w_1\rangle\in\rho(\up)$ (or $\langle
w,w_1\rangle\in\rho(\dx)$)
and
$\langle w,w_2\rangle\in\rho(\up)$ (or $\langle
w,w_2\rangle\in\rho(\dx)$),
}
$w\Up w_1$ (or $w\Dx w_1$) and $w\Up w_2$ (or $w\Dx w_2$), 
then 
$w_1=w_2$.
Consequently, 
total functions $up:Grid\imp Grid$ and $right:Grid\imp Grid$
can be defined as follows:
\begin{itemize}
\item for all $w\in Grid$, $up(w)=w'$ iff 
$w\Up w'$;
\item for all $w\in Grid$, $right(w)=w'$ iff 
$w\Dx w'$.
\end{itemize}

$\M_{w_0}\models \gamma$ implies that for every $w\in Grid$,
there exist (unique states, by $\beta$) $w_1,w_2,w_3$  such that
$w \Up w_1\Dx w_2$ and $w\Dx w_3\Up w_2$.
Therefore:
\begin{itemize}
\item[(1)] for all $w\in Grid$, $right(up(w))=up(right(w))$.
\end{itemize}
Actually, in the presence of $\beta$, $\gamma$ is equivalent to:
\[\Box_{\grid}\binder x.\Box_{\up}\Box_{\dx}\binder y.
x:\Box_{\dx}\Box_{\up} y\]

Let $f:\Nat\times\Nat\imp Grid$ be defined as follows:
$f(0,0)=w_0$, $f(n+1,m)=up(f(n,m))$, 
and $f(n,m+1)=right(f(n,m))$. This function is well
defined, because $up$ and $right$ are well defined, and,
moreover, $f(n+1,m+1)$ is uniquely defined, since:
\[\begin{array}{l}
f(n+1,m+1)=rigth(f(n+1,m))=up(right(f(n,m)))\\
f(n+1,m+1)=up(f(n,m+1))=right(up(f(n,m)))
\end{array}\]
and $up(right(f(n,m)))=right(up(f(n,m)))$ by (1).

The function  $tile:\Nat\times\Nat\imp T$ is then defined as follows:
$tile(n,m)=t_i\in T$ iff $\M_{f(n,m)}\models p_{t_i}$.
Using the fact that $\M_{w_0}\models \delta$, it easily follows that
$tile$ is a tiling of $\Nat\times\Nat$.

For the converse, we show how to build a model of $\pi_T$ from a
tiling $tile:\Nat\times\Nat\imp T$ of $\Nat\times\Nat$.
\[\begin{array}{l}
W=\{(n,m)\mid n,m\in\Nat\}\\
I(a)=(0,0)\\
\Grid
=\{\langle (0,0),(n,m)\rangle ,\langle (n,m),(0,0)\rangle \mid  n,m\in\Nat\}\\
\Up=\{\langle (n,m),(n,m+1)\rangle \mid  n,m\in\Nat\}\\
\Dx=\{\langle (n,m),(n+1,m)\rangle \mid  n,m\in\Nat\}
\end{array}
\]
By construction, $\M_{(0,0)}\models \pi_T$.

The fact that the
satisfiability problem for
$\Hl(@,\binder,\Diamond^1)\setminus \Box\binder\Box$ is undecidable
follows directly from the undecidability of the 
$\Nat\times\Nat$ tiling problem, since  $\pi_T$ 
does not contain
the pattern $\Box\binder\Box$.
\end{proof}

The above result might be of poor interest in the
context of description logics, since their language does not have
full use of the satisfaction operator.  However,
its use in the encoding of the tiling problem can be replaced by
the  converse modalities.
 The hybrid multi-modal language with binders, converse modalities and
both forward and backward functional restrictions will be denoted by
 $\Hl_m(\binder\comma\Diamond^-\comma\Diamond^1)$, and its fragment
consisting of formulae 
whose NNF
do 
not contain any occurrence of the binder
that is both in the scope and has in its scope a universal modality
({i.e.} either $\Box_\R$ or 
$\Box_\R^1$)  is denoted by 
$\Hl_m(\binder\comma\Diamond^-\comma\Diamond^1)\setminus \Box\binder
  \Box$.

\begin{theorem}
The satisfiability problem for 
$\Hl_m(\binder,\Diamond^-,\Diamond^1)\setminus \Box\binder
  \Box$
is undecidable.
\end{theorem}

\begin{proof}
The $\Nat\times\Nat$ tiling problem can be encoded in 
$\Hl_m(\binder,\Diamond^-,\Diamond^1)\setminus \Box\binder
  \Box$, by use of the same formulae $\alpha,\beta$ and $\delta$ used
  in the proof of Theorem \ref{undec-at}, and the following ones,
  replacing $\gamma$:

\begin{description}

\item[Injectivity.]
$\beta' = \Box_{\grid}\Box_{\up^-}^1 \bot \wedge 
\Box_{\grid}\Box_{\dx^-}^1 \bot$.
\item[Grid.] $\gamma^- = \Box_{\grid}\Box_{\up}\Box_{\dx}\binder
x.\Diamond_{\dx^-}\Diamond_{\up^-}\Diamond_{\dx}\Diamond_{\up} x$.
 \end{description}

$\beta'$ states that every state in the grid has 
at most one $\Up$ predecessor and
at most one $\Dx$ predecessor.
And $\gamma^-$  states that for
 every state $x$ of the grid
  that can be accessed from some state going up ($\Up$) and
then right ($\Dx$),  it is possible to move from $x$ to $x$
itself, by moving left ($\Dx^-$),
 then down ($\Up^-$),  then right, and then up.

In order to show that any set of tile types $T$ tiles $\Nat\times\Nat$ iff
    the  formula $\pi^-_T =
    \alpha\wedge\beta\wedge\beta'\wedge\gamma^-\wedge\delta$
 is satisfiable, the proof of Theorem \ref{undec-at}
must be slightly modified to show that, in the presence of the other
formulae, $\gamma^-$ expresses the desired grid property.

Since $\M_{w_0}\models\beta'$, for every $w,w_1,w_2\in Grid$, if
$w_1\Up w$ (or $w_1\Dx w$) and $w_2\Up w$ (or $w_2\Dx w$), then 
$w_1=w_2$.  Consequently, the functions $up$ and $right$, defined like
in the proof of Theorem \ref{undec-at},
are injective, and their converses, $down=up^{-1}$ and $left=right^{-1}$ are 
well defined partial functions on $Grid$.

$\M_{w_0}\models \gamma^-$ implies that for every $w,w_1,w_2\in Grid$,
if $w_1\Up w_2\Dx w$, then also $w_1\Dx w_3\Up w$ 
for some $w_3\in Grid$. In fact, if
$w_1\Up w_2\Dx w$  ({\ie} $w=right(up(w_1)$)
and $\sigma(x)=w$, then
\[\M_{w},\sigma\models
\Diamond_{\dx^-}\Diamond_{\up^-}\Diamond_{\dx}\Diamond_{\up} x\] 
Since   $\M_{w_0}\models\beta'$, $w_1$ and $w_2$ are the unique states such that
 $w_1\Up w_2\Dx w$, {\ie} $w_1=left(down(w))$.
Therefore
\[\M_{w_1},\sigma\models\Diamond_{\dx}\Diamond_{\up} x\] 
In other terms, the (unique, by $\beta$) state $w'$ such that 
$w_1\Dx w_3\Up w'$ is $w$, {\ie} $w=up(right(w_1))$.  So:
\begin{itemize}
\item[(1)] for all $w\in Grid$, $right(up(w))=up(right(w))$.
\end{itemize}
This is enough to exploit the rest of the proof of Theorem
\ref{undec-at}.
\end{proof}

%% file: preprocessing.tex
\section{The preprocessing step of the satisfiability decision
  procedure}
\label{preprocess}

Let $F$ be a formula in NNF belonging to the fragment
 $\mmframmento$ and $\calA$ a set
of assertions. 
In order to test $\{F\}\cup\calA$ for satisfiability
by means of the calculus presented in Section \ref{calcolo},
$F$ is first preprocessed and translated into an equisatisfiable
formula in the fragment $\Hl_m(@\comma \binder\comma \E\comma \Diamond^-)
  \setminus\binder\Box$, {\ie} a formula in NNF where no universal
  operator (either $\Box_R$ or $\A$) occurs in the scope of a
  binder.\footnote{Graded modalities may also be allowed in the input
    formula $F$, provided they satisfy
the restrictions  stated in Theorem \ref{graded-dec}. If this is the
case,
 they are 
eliminated beforehand, and replaced by their definitions in 
$\Hl_m(@\comma \binder\comma \E\comma \Diamond^-)\setminus\Box\binder\Box$,
 as shown in the
proof of the above mentioned theorem.}
\margin{footnote modif} 
\butta{sulla footnote 1: I don't get the point about this
footnote, since the language considered doesn't even contain graded
modalities.
}

The translation is the multi-modal analogous of the (polynomial)
satisfiability preserving translation given in \cite{FranBis} for
 $\frammento$. 
Its 
definition is given below  in
order to make the paper self contained.
It 
 differs from the way it is defined in \cite{FranBis}, though actually
 equivalent.\footnote{The translation given in \cite{FranBis} (Theorem 1)
preserves satisfiability, but the given justification of this fact
is incorrect. 
In fact, it uses, among others,  the ``equivalence'' $t\at\exists x F \equiv
\exists x \,t\at F$  (where $\M_w, \sigma \models\exists x\, G$ iff 
$\M_{w}, \sigma^{w'}_x \models G$ for some state $w'$) that does not
hold, as
 it can easily be seen by
considering $a\at \exists x (x \wedge x\at  \neg a)$ and
$\exists x \,a\at  (x \wedge x\at  \neg a)$.}

\butta{
\margin{modif} Dire nella nota che anche se definito diversamente, il
risultato della traduzione e' lo stesso di quello di ten cate e
franceschet.  the definition embeds a proposition: the fact that the
translation preserves satisfiability. This claim should not be part of
the definition, but should be a lemma/proposition. A proof of it would
also be nice, in particular since you claim that the proof of
satisfiability preservation of the translation given in [25] is
incorrect, and that you use a different translation!
}

\margin{vorrebbe la dimo di equisat}
\begin{definition} Let $F$ be a formula in NNF 
in $\mmframmento$.
The  translation $\tau(F)$ of $F$
 into an equisatisfiable formula in 
$\Hl_m(@\comma \binder\comma \E\comma \Diamond^-)
  \setminus\binder\Box$ is inductively defined as follows:
\[\begin{array}{rclp{10em}}
\tau(u\at F) &~=~& u\at \tau(F) \mbox{~~~ where }u\in\NOM\cup\X\\
\tau(F\wedge G) &=& \tau(F)\wedge \tau(G)\\
\tau(F\vee G) &=& \tau(F)\vee \tau(G)\\
\tau(\Diamond_\R F) &=& \Diamond_\R \tau(F)\\
\tau(\E F)  &=& \E \,\tau(F)\\
\tau(\binder x.F) &= &
\left\{
\begin{array}{lp{20em}}
\binder x. F & ~if $F$ contains no universal operator\\
b\wedge \tau(F[b/x]) & ~(where $b$ is a fresh nominal) otherwise
\end{array}\right.\\
\tau(F) &=& F  \mbox{~~~ in all the other cases}
\end{array}\]
\end{definition}

Above, fresh nominals are nominals that not only do not occur in the
formula to be translated, but also are  used nowhere else in the translation.
For instance:
\[\begin{array}{ll}
\tau((\A\binder x.\Diamond_r x) \wedge
((\binder y.\Box_r y)\vee(\binder z.\A z)))\\
=  (\A \binder  x.\Diamond_{r}x)     \wedge  
((a_{1}  \wedge  \Box_{r}a_{1}  )  \vee  (a_{2}  \wedge  
\A a_{2}  ) )
\end{array}\]

Assuming that $F$ does not contain the pattern $\Box\binder\Box$, and 
 $\binder x.G$ is a a subformula of $F$, if $G$ contains a universal
operator, then $\binder x.G$ does not occur in the scope of a binder
in $F$. Therefore $\tau(\binder x.G)$  is a kind of skolemization inside
$F$ of 
 $\binder x.G$. If on the contrary $G$ does not contain universal
operators, then the subformula $\binder x.G$ cannot be responsible of
the critical pattern in $F$ and is left unchanged.

%% file: calcolo.tex
\section{The tableau calculus}
\label{calcolo}

\butta{
\margin{modif} In the beginning of Section 5 you ought to state for
which language it is you provide a set of tableau rules.
}

This section shows how to extend the system described in
\cite{jar2012} to the presence of transitivity and inclusion
assertions, obtaining a tableau calculus for 
$\Hl_m(@,\binder,\E,\Diamond^-,\trans,\subrole)$.
  The expansion rules that will be introduced to treat
assertions are similar to the analogous ones presented in
\cite{HoSa99,HoSa07a,KS-graded-role-hierarchies,KaSmoJoLLI2009}.
However, their addition to a terminating calculus dealing also with
syntactically restricted occurrences of the binder is a novelty.

The presentation will be as self contained as possible, therefore it
overlaps with the description given in \cite{jar2012} in many points.
However, since some of the basic notions underlying the calculus are
quite involved, they are not given a completely formal account here,
but are rather taken as an opportunity to explain some subtle notions
in  intuitive terms.

\subsection{The calculus}

 A tableau is a set of branches, and a {\em tableau branch} is a
 sequence of {\em nodes} $n_0,n_1,\dots $, where each node is labelled
 either by an assertion or a ground {\em satisfaction statement},
 {\ie} a formula of the form $a\at F$, where no state variable occurs
 free in $F$. The nominal $a$ in a satisfaction statement $a\at F$ is
 called the {\em outermost nominal} of the formula and $F$ its {\em
   body}.  Node labels are always formulae in NNF.  
\butta{
The reason why a
 branch is not simply a set of formulae will be briefly explained in
 the sequel.
}

Statements of the form $a\at\Diamond_\r b$, where $a$ and $b$ are
nominals and $\r$ is a forward relation are called {\em relational
  formulae}, and nodes labelled by relational formulae are called {\em
  relational nodes}.  Expressions of the form $a\Rightarrow_{\R}b$
will be used as abbreviations for  relational formulae:
\[\begin{array}{lcl}
a\Rightarrow_{\R}b & \equiv_{def} &
      \left\{\begin{array}{ll}
             a\at\Diamond_\r b & \mbox{if }{\R}=\r\\
             b\at\Diamond_\r a & \mbox{if }{\R}=\r^-
      \end{array}\right.
\end{array}
\]
By convention, an expression of the form  $\trans(\R)$, where $\R$ is
a meta-symbol standing for either a forward or backward relation, will
stand for $\trans(\r)$, where $\r\in\REL$ is the relation symbol in $\R$.

  If $n$ occurs before $m$ in a branch, we write $n<m$.  The label of
  the node $n$ is denoted by $\operatorname{label}(n)$.  The notation
  $(n)\,a\at F$ is used to denote the node $n$, and simultaneously say
  that its label is $a\at F$.  If a node $(n)\,a:F$ is in a branch,
  then the nominal $a$ is said to label the formula $F$ in the branch.

Let $F$ be a ground hybrid formula in NNF and $\calA$ a set of
assertions.  A tableau for $\{F\}\cup\calA$ is initialized with a
single branch, constituted by the node $(n_0)\,a_0\at F$, where $a_0$
is a new nominal, followed by nodes labelled by the assertions in
$\calA$ and then expanded according to the {\em Assertion rules} of
Table \ref{assertion-rules} (note that $\rel$ actually stands for four
rules, according to the relation signs).  Such rules complete the
inclusion assertions in $\calA$ by the reflexive and transitive
closure of $\subrole$.  The formula $a_0\at F$ is the {\em initial
  formula} of the tableau.

\begin{table}[htb]
{\footnotesize
\[
\begin{array}{|ccc|}
\hline
&&\\
~~~~\infer[\rel_0]{\r\subrole \r}{}
&~~~
& \infer[\rel]{\R\subrole \P}{\R\subrole \S & \S\subrole\P}~~~\\
&&\\
\hline
\end{array}
\]}
\caption{Assertion rules}
\label{assertion-rules}
\end{table}

A tableau branch is expanded by either adding nodes or changing node
labels, according to  the rules in Table
 \ref{regole}. 
Most rules are standard, and their reading is standard too.
Note that when the formulation of a rule contains (uppercase)
relations, it actually stands for different rules, according to the
relations signs.  
In applications of either the $\Diamond$ or the $\E$ rule, the nominal
$b$ occurring in the conclusion(s) is fresh in the branch.
Moreover, the $\Diamond$ rule is not applicable to relational nodes
(where $\R$ is a forward
relation and $F$ is a nominal).  In applications of the $\A$ rule, the
nominal $b$ is any nominal occurring in the branch.
 The {\em equality rule} $(=)$ does not
add any node to the branch, but modifies the labels of its nodes. The
schematic formulation of this rule in Table \ref{regole}
indicates that it can be fired whenever a branch $\ramo$ contains a
{\em nominal equality} of the form $a\at b$ (with $a\neq b$); as a
result of the application of the rule, every node label $F$ in $\ramo$
is replaced by $F[b/a]$.

\begin{table}[htb]
{\footnotesize
\[
\begin{array}{|cc|}
\hline
&\\
\infer[(\wedge)]{\begin{array}{c}(m_0)\,a\at F\\ (m_1)\,a\at G\end{array}}
{(n)\,a\at (F \wedge G)} &
\infer[(\vee)]{(m_0)\,a\at  F~~~\mid~~~(m_1)\,a\at  G}
{(n)\,a\at (F \vee G)}~~~
  \\
&\\
\infer[(@)]{(m)\,b\at F}{(n)\,a\at  b\at F}
&
\infer[(\binder)]{(m)\,a\at F[a/x]}
    {(n)\,a\at \binder x.F} 
\\
 &\\
~~~{\infer[(\Box)]
{(k)\,b\at F}{(n)\,a\at\Box_{\R}F & (m)\,a\Rightarrow_{\R} b}}
&
\infer[(\Diamond )]
{\begin{array}{c}(m_0)\,a\Rightarrow_{\R} b\\(m_1)\,b\at F\end{array}}
{(n)\,a\at \Diamond_\R  F}
\\
&\\
\infer[(\A)]{(m)\,b\at F}{(n)\,a\at \A F} &
\infer[(\E)]{(m)\,b\at F}{(n)\,a\at \E F}\\
&\\
\infer[(=)] {{\cal B}[b/a]}
{\begin{array}{c}[{\cal B}]\\ (n)\,a\at b\end{array}}
 &
\infer[(\link)]{(m)\,a\Rightarrow_{\S} b}
{(n)\,a\Rightarrow_{\R}b &  (i)\,\R\subrole {\S} }
\\ &\\
\multicolumn{2}{|c|}{\infer[(\trans)]{(k)\,b\at \Box_{\R} F}
{(n)\,a\at \Box_{\S} F &
(m)\,a\Rightarrow_{\R}b & (t)\,\trans({\R}) & 
(i){\R}\subrole {\S}}}\\
&\\
\hline
\end{array}
\]}
\caption{Expansion rules}
\label{regole}
\end{table}

Formulae of the form $\Box_{\R}F$ and $\A F$ are called
{\em universal formulae}; nodes whose labels have the form $a\at G$,
where $G$ is a universal formula, are
{\em universal nodes} and the rules $\Box$ and $\A$ are called {\em
  universal rules}.  When the $\A$ rule is applied producing a node
labelled by a formula of the form $b:F$, it is said to {\em focus} on
$b$ (and $b$ is the focused nominal of the inference).  The
$\Diamond$ and $\E$ rules are called {\em blockable
  rules}, non relational 
formulae of the form $a\at \Diamond_{\R} F$
and $a\at\E F$ are {\em blockable
  formulae} and a node labelled by a blockable formula is a {\em
  blockable node}. 
The $\trans$ rule  deals with transitive
relations and can be seen as a reformulation (in the presence of
inclusion assertions) of the $\Box$ rule for transitive modal logics
(a particular case of this rule is when $\R=\S$).

The premiss $n$ of either the $\Box$ or $\trans$ rules is called the
{\em major premiss}, and $m$ the {\em minor premiss} of the rule.  In
an application of the $\link$ rule, $n$ is its {\em logical premiss}.


The first node of a branch $\ramo$ is called the {\em top node} and
its label the {\em top formula} of $\ramo$.  Nominals occurring in the
top formula are called {\em top nominals}.  The notion of top nominal
is relative to a tableau branch, because applications of the equality
rule may change the top formula, hence the set of top nominals.

A branch is {\em closed} whenever it contains, for some nominal $a$,
either a pair of nodes $(n)\, a\at p$, $(m)\,a\at \neg p$ for some
$p\in\PROP$, or a node $(n)\,a\at \neg a$.  As usual, it is assumed that a
closed branch is never expanded further.  A branch which is not
closed is {\it open}.  A branch is {\em complete} when it cannot be
further expanded.

Provided that the initial formula is in $\Hl_m(@\comma \binder\comma
\E\comma \Diamond^-) \setminus\binder\Box$, the calculus enjoys the
following important {\em strong subformula property}, that is essential to prove
both termination and completeness:  every universal formula
occurring in a tableau branch is obtained from a subformula of
the top formula $F_0$ of the branch by possibly replacing  
operators $\Box_\R$
with $\Box_\S$, for some relation $\S$ in the language of the initial tableau
(see Lemma \ref{subformula} in the Appendix).
\butta{
Treating nominal equalities by means of substitution, like in
\cite{CeCi09a,aiml-nostro,jar2012,herod-ijcar}, is essential to ensure
such a property:
if some kind of ``copy rule'' were used instead,
any complete branch containing an equality $a\at b$
 and a node labelled by a formula
of the form $c\at\Box_{\R}F(a)$, 
would contain also $c\at\Box_{\R}F(b)$, and, in general, $F(b )$ cannot be
ensured to be a subformula of the top formula.

\margin{modif} "Treating nominal equalities by means of substitution,
like in [5, 6, 8, 10], is essential to ensure such a property: ...
 That certainly depends
on what kind of "copy rule" you choose. For basic hybrid logic, it is
sufficient to have a copy rule that allows you to conclude from
premises a:b and a:F to conclusion b:F. So I don't think you have a
strong argument here about substitution being essential for the strong
subformula property. Either you have to support your claim by a
stronger argument, or simply delete the claim (it is not essential for
the paper).
}

By the effect of substitution, however, distinct node
labels may become equal, though the corresponding nodes are still
distinct elements of the branch.
The reason why nodes with the same
label do not collapse and a branch is not simply a set of formulae
 is explained in next subsection.
\butta{
is that they must be arrangeable in a tree-like
structure, where each node has at most one parent. 
Next sub-section introduces the 
 relation on
nodes inducing such a structure and the related notion of blocking.
}

\subsection{Blocking and other restrictions on rule application}

\margin{ Please state a reference to the source of the idea of
direct blocking}

Termination is achieved by means of a form of anywhere blocking with
indirect blocking.  Direct blocking must take into account the fact
that, due to the presence of the binder, a potentially infinite number
of distinct nominals may occur in the bodies of node labels (the
strong subformula property only holds for universal formulae).  The
$\Diamond$ and $\E$ expansion rules, in fact, add fresh nominals to
the branch and the expansion of a node $(n)\,a\at\binder x.F$ produces
a node containing $a$ in the body of its label.  As a consequence, a
branch may contain an infinite number of blockable formulae pairwise
differing not only for the respective outermost nominals.

Mainly, direct blocking is a relation between nodes in a tableau
branch, holding whenever the respective labels (formulae) are equal up
to (a proper form of) nominal renaming.  Essentially, in order for a
node $(n)\,F$ to (directly) block $(m)\,G$ in a branch $\ramo$, it
must be the case that $G=F[a_1/b_1,\dots,a_n/b_n]$, where
$a_1\comma\dots,a_n\comma b_1\dots\comma b_n$ are {\em non-top}
nominals such that, for all $i=1,\dots,n$, $a_i$ and $b_i$ label the
same set of formulae of a certain kind in $\ramo$.
The (direct) blocking restriction forbids the application
of a blockable rule to a node $n$, whenever the label of a node $m<n$
can be mapped in that way to $\operatorname{label}(n)$. 

The precise definition of direct blocking is given later on
(Definition \ref{blocking}). What is important to point out here is
that, differently from other tableau calculi for $\Hl$, blocking is a
relation between nodes, not nominals.

When a node is blocked, all its {\em descendants} {\wrt} a particular
relation called the {\em offspring relation} are indirectly blocked
and are called {\em phantom nodes}.  The offspring relation, denoted
by $\freccione_{\cal B}$, is a partial order arranging 
the nodes of a branch into a
tree-like structure, where each node has at most one parent and non-terminal
nodes are blockable nodes.  Every tree is rooted at a node called a
{\em root node} (a node with no {\em parents} {\wrt} the offspring
relation). 

The nodes of the initial tableau are all root nodes.  Blockable rules
generate {\em children} ({\wrt} the offspring relation) of the
expanded node: if the expansion of a blockable node $n$ generates
$m_0$ (and $m_1$), then $n\prec_{\cal B}m_0$ (and $n\prec_{\cal B}m_1$).
  All the other rules, with the exception of the $\A$
rule, generate {\em siblings} of one of the premisses of the
inference (two nodes are called siblings 
 if either they are both root nodes or they have the same parent).
 For instance, if a node $n$ is expanded by means of
the $\wedge$ rule generating $m_0$ and $m_1$, then if $n$ is a root
node then also $m_0$ and
$m_1$ are root nodes; otherwise, if $k\prec_{\cal B} n$ for some node
$k$, then also $k\prec_{\cal B} m_i$ ($i=0,1$).

The offspring relation is formally defined in Definition
\ref{offspring}. In order to understand its behaviour {\wrt}
rules applied to universal nodes, is important to mention one of the
important properties on which the termination proof relies, {\ie} that
any node has a bounded number of siblings (Lemma \ref{base} in the
Appendix).  In order to prove such a property, it is essential that,
when the $\Box$ rule expands a pair of nodes $(n)\,a\at\Box_\R F$ and
$(m)\,a\Rightarrow_\R b$, the conclusion $k$ is a sibling of the {\em
  minor premiss} $m$ of the inference.  If, on the contrary, $k$ were
a sibling of $n$, then $n$ might have an infinite number of siblings,
since, in principle, there might be an unbounded number of nominals
$b_i$ such that $a\Rightarrow_\R b_i$ is in the branch.  Similarly, a
node obtained by use of the $\trans$ rule is a sibling of the minor
premiss of the inference.

The $\A$ rule is however problematic, since it
can also be applied several times to the same node generating a
potentially unbounded number of different
conclusions. These nodes cannot be siblings of the premiss, 
that, otherwise, could have an unbounded  number
of siblings.  Analogously to the $\Box$ rule, the
$\A$ rule needs a minor
premiss, to be taken as a sibling of the conclusion. It is then
established that 
the minor premiss  of an
application of the $\A$ rule is the first {\em non-phantom} node where
the focused nominal $b$ occurs, in the branch where the rule is
applied
 (termination relies also on the fact that 
phantom nodes cannot be used as minor premisses of any
rule -- see Definition \ref{restriction}).

Apparently, there is a circularity in this definition: phantom nodes
are defined in terms of $\prec_{\cal B}$, which is in turn defined
assuming to know which nodes are phantoms.
Properly, the offspring relation and blockings are defined
contemporarily by induction on  branch construction: 
\begin{itemize}
\item in the initial tableau no node is blocked and all nodes are root
  nodes;
\item let us assume that the set of (directly and indirectly) blocked nodes
  of a branch $\ramo$ is defined, and that $\ramo$ is expanded to
  $\ramo'$; then
the offspring relation in $\ramo'$ is defined in terms of the phantom
nodes in $\ramo$, and $\prec_{{\cal B}'}$ is used (together with
direct blocks in $\ramo'$) to determine which nodes are phantoms in
$\ramo'$.
\end{itemize}
The presentation that follows is somewhat simplified, and the reader
is referred to \cite{jar2012} for the more formal approach.  

\begin{definition}
\label{minor-premiss}
Let $\ramo'$ be obtained from $\ramo$ by means of an
application $\cal I$ of the $\A$ rule focusing on the nominal $b$, and
let us assume that the set of phantom nodes in $\ramo$ is already defined.
Then the
minor premiss of $\cal I$ 
is
the first non-phantom node in $\ramo'$ where $b$ occurs.
\end{definition}
\noindent
Note that
in principle, an application of the $\A$ rule could have no minor
premiss (when the focused nominal only occurs in phantom nodes). This
possibility, however, will be ruled out by the restrictions on rule
applications that are introduced later on.

Knowing which are the minor premisses of applications of the $\A$ rule in
a branch $\ramo$, the offspring relation $\prec_{\cal B}$ can be
defined. It is a static relation: if $n\prec_{\cal B}m$ and $\ramo'$
is obtained as an expansion of $\ramo$, then also 
 $n\prec_{{\cal B}'}m$.  Hence, in particular,  the top node and
all the nodes labelled by assertions are root nodes in any branch.

\begin{definition}[Offspring relation]
\label{offspring}
Let $\ramo$ be a tableau branch, and let $\ramo'$ be an expansion of
$\ramo$.
Then:
\begin{enumerate}
\item  if $n\prec_{\cal B}m$, then also 
 $n\prec_{{\cal B}'}m$. 
\item If $\ramo'$ is obtained from $\ramo$ by application of a
  blockable rule to a node $n$, adding the new node(s) $m_0$  (and
  $m_1$), then $n\prec_{{\cal B}'}m_i$ ($i=0,1$).
\item 
\label{universali}
If $\ramo'$ is obtained from $\ramo$ by application of  either a universal
  rule or the $\trans$ rule whose minor premiss is $m$, 
adding the new node $k$, 
then $k$ is a sibling of $m$
  (i.e., if $m$ is a root node, then $k$ is a root node too;
  otherwise, if $k'\freccione_{\cal B} m$, then 
$k'\freccione_{\cal B} k$).
\item If $\ramo'$ is obtained from $\ramo$ by application of the
  $\link$ rule, then the newly added node is a sibling of the logical
  premiss of the inference.
\item If $n$ $\ramo'$ is obtained from $\ramo$ by application of any
  other rule of Table \ref{regole} which adds new nodes  ({i.e.} any
  other single-premiss 
  rule, excluding the equality rule), then the conclusions are
  siblings of the premiss of the rule application.
\end{enumerate}
\end{definition}

As it has already been pointed out, the termination proof essentially
relies on the fact that the offspring relation arranges the nodes of a
branch into a bounded sized set of trees, each of which has bounded
width (and bounded depth -- which will be ensured by blocking).  
This holds because a branch is not a set
of formulae, but nodes, and each node has at most one parent.  If
nodes labelled by the same formula collapsed into a single branch
element, such an element might have multiple parents. For a similar
reason it is not possible to block nominals instead of nodes: two
nominals with different ``parents'' may become equal by substitution.

The drawback is that the reasoning proving that any node has a bounded
number of siblings is not as simple as it would be if dealing with
sets of formulae. It relies in an essential way on the fact that
universal rules do not generate siblings of their major premisses and,
thanks to the already mentioned strong subformula property,
the number of universal formulae occurring in a tableau branch is bounded.


Once the offspring relation has been introduced,
the notions of direct and indirect blocking can be formally
defined, preceded by the conditions on nominal renaming required
for a formula to be ``mappable'' to another one. 

\begin{definition}[Nominal compatibility and mappings]
\label{compatib}\label{mappings}
If {\ramo} is a tableau branch, then:
\begin{enumerate}

\item two nominals $a$ and $b$ are {\em compatible} in {\ramo} if they
  label the same propositions in $\PROP$ and the same formulae of the
  form $\Box_{\R} F$, {\ie}
\[\begin{array}{rll}
\{p\in\PROP\mid a\at p\in\ramo\} & = & \{p\in\PROP\mid b\at
p\in\ramo\}\\
\{\Box_{\R}F\mid  a\at \Box_{\R}F\in\ramo\} & = &
\{\Box_{\R}F\mid  b\at \Box_{\R}F\in\ramo\}
\end{array}
\]
(where $G\in\ramo$ stands for ``$G$ is the label of some node in
$\ramo$'').

\item 
A mapping $\pi$ for $\ramo$ is an {\em injective} function from {\em
  non-top} nominals to {\em non-top} nominals such that for all $a$,
$a$ and $\pi(a)$ are compatible in \ramo.  Mappings are extended to
act on formulae in the obvious way: $\pi(F)$ is the formula obtained
by substituting $\pi(a)$ for $a$ in $F$, for every non-top nominal $a$.

\item
A mapping $\pi$ for $\ramo$ maps a formula $F$ to a formula $G$ if
$\pi(F)=G$ and $\pi$ is the identity for all nominals which do not
occur in $F$.

\item
A formula $F$ can be mapped to a formula $G$ in $\ramo$ if there
exists a mapping $\pi$ for $\ramo$ mapping $F$ to $G$.
\end{enumerate}
\end{definition}

\begin{definition}
[Direct and indirect blocking]
\label{blocking}
 Let $\ramo$ be a tableau branch.  The
set of directly and indirectly blocked nodes in $\ramo$ is defined by
induction on the (total) order $<$ on the nodes of $\ramo$:
\begin{itemize}
\item $n$ is blocked if it is either directly or indirectly blocked.
\item 
$n$ is directly blocked by $m$ if $n$ is a blockable node, $m<n$, $m$
  is not blocked and $\operatorname{label}(m)$ can be mapped to
  $\operatorname{label}(n)$ in \ramo; $n$ is directly blocked in
  $\ramo$ if it is directly blocked by some $m$ in $\ramo$.
\item $n$ is indirectly blocked if it is not directly blocked and it
  has an ancestor {{\wrt}} $\freccione_{\cal B}$ which is blocked.
\end{itemize}
An indirectly blocked  node  is called a {\em phantom node} (or,
simply, a phantom).
\end{definition}
It is worth noticing that  blocked nodes are not required to be
$\prec_{\cal B}$ descendants of the respective blockers, and that 
a node is a phantom if and only if all its
siblings are phantoms too.

Blockings induce the following restrictions on branch expansion:

\begin{definition}[Restrictions on the expansion rules]
\label{restriction}
The expansion of  a tableau branch {\ramo} is
subject to the following restrictions:
\begin{description}
\item [{\bf R1.}] no node labelled by a formula already occurring in
  {\ramo} as the label of a {\em non-phantom} node is ever added to
  \ramo.

\item [{\bf R2.}]  Blockable nodes can be expanded at most once in a
  branch. 

\item[{\bf R3.}]  
A phantom node cannot be expanded by means of a
  single-premiss rule (including the equality rule), 
it cannot be used as the
 logical premiss of an application of the $\link$ rule,
nor can it be
  used as the minor premiss of a universal rule or the $\trans$
  rule.

\item[{\bf R4.}] A blockable node $n$ cannot be expanded if it is
  directly blocked in $\ramo$.

\butta{
\item[{\bf R5.}] A phantom node cannot be  used as the minor premiss of
   the $\trans$
  rule.
\item[{\bf R6.}] A phantom node cannot be  used as the
 logical premiss of the $\link$ rule.
}
\end{description}
\end{definition}

Note that, as a particular case of restriction {\bf R3},
the $\A$ rule cannot 
focus on a nominal which only occurs in phantom nodes in the branch.
Consequently, thanks to this restriction, every application of the
$\A$ rule has a minor premiss.
\medskip

Termination and completeness are stated and proved in detail in the
Appendix. 
It is worth pointing out here that, according to the termination
proof, 
the worst-case complexity of the
calculus presented in this work has the same order of magnitude of the
calculus in \cite{jar2012}: the termination proof given in the 
Appendix shows that
the nodes of a tableau branch are arranged
by $\freccione_{\cal B}$ in a forest of trees, whose number is bounded
by an exponential function of the size $N$ of the input problem.
 Both
tree width and tree depth are bounded by exponential functions of $N$, 
 therefore the
number of nodes in a single branch is bounded by a doubly exponential
function. \butta{
 Therefore, the number of nodes in a tableau is bounded by a
triple exponential function of $N$.
As a consequence, according to the bounds given above, the
decision procedure defined in this paper is not worst-case optimal,
since the satisfiability problem for $\fhl\,\setminus \binder\Box$ is
in {\sc 2ExpTime} \cite{FranBis}.
}
Since the cost of blockings is in the order of the branch size, 
 the tableau
calculus presented in this work shows that the satisfiability problem
for $\thframmento$ is in {\sc 2-NExpTime}.  It is reasonable to
hypothesize that the problem complexity is actually lower,%
\footnote{%
The satisfiability problem for $\miniframmento$ is in {\sc
  2ExpTime} \cite{FranBis}, and the complexity of the concept
satisfiability problem in description logics does not increase with
the addition of transitive roles and role hierarchies.  } 
and consequently
  that, like many other tableau based algorithms, the decision
  procedure defined in this paper is not worst-case
  optimal.


%% file: examples.tex
\section{Examples}
\label{examples}

This section contains some examples illustrating the calculus in
action.\footnote{The second example (Figure 1)
 in \cite{cade2013} is incorrect. An {\em
    Errata corrige} is available at the author's web page.}
In the tableau represented below, 
the notations $n\leadsto^{\cal
  R} m$ or $(n_1,\dots,n_k)\leadsto^{\cal R} m$
 mean that the addition of node $m$ is due to the
application of rule $\cal R$ to node $n$ (or nodes $n_1,\dots,n_k$).
If ${\cal R}=\A$, then also the minor premiss is  indicated:
$(n,m)\leadstoA k$ means that the $\A$ rule is applied to $n$ with
minor premiss $m$, producing $k$.
Moreover, 
the notation $n\prec_{\cal B}
  \{m_1,\dots,m_k\}$, used to illustrate the offspring relation,
 abbreviates $n\prec_{\cal B} m_1$ and \dots
  $n\prec_{\cal B} m_k$.

\begin{example}
The  simple example represented in Figure \ref{ese1}
 shows the interplay between the
$\trans$ and $\link$ rules.  
It consists of a closed one-branch
tableau  for the formula $\Diamond_{\s}\Diamond_{\s} p
\wedge \Box_{\s}\neg p$, together with the assertions
$\trans(\r),~\r\subrole \s,~\s\subrole \r$. The branch is closed
because of nodes $11$ and $15$.
In this branch, $0$--$7$ are root nodes,
$6\prec_{\cal B}\{8,9,12,14\}$, and
$9\prec_{\cal B}\{10,11,13,15\}$.

\begin{figure}[htb]
{
\[\begin{array}{llc}
\begin{array}{rll}
(0) & a\at(\Diamond_{\s}\Diamond_{\s} p \wedge \Box_{\s}\neg p)\\
(1) & \trans(\r)\\
(2) & \r\subrole \s\\
(3) & \s\subrole \r\\
(4) & \r\subrole \r & \rel_0\\
(5) & \s\subrole \s & \rel_0\\
(6) & a\at\Diamond_{\s}\Diamond_{\s} p & 0\leadsto^\wedge 6\\
(7) & a\at\Box_{\s}\neg p& 0\leadsto^\wedge 7\\
\end{array}
&~~~~~~&
\begin{array}{rlc}
(8) & a\at\Diamond_{\s} b & 6\leadsto^\Diamond 8\\
(9) & b\at \Diamond_{\s} p& 6\leadsto^\Diamond 9\\
(10) & b\at\Diamond_{\s} c & 9\leadsto^\Diamond 10\\
(11) & c\at p &  9\leadsto^\Diamond 11\\
(12) & a\at\Diamond_{\r} b & (8,3)\leadsto^{\link} 12\\
(13) & b\at\Diamond_{\r} c & (10,3)\leadsto^{\link} 13\\
(14) & b\at\Box_{\r} \neg p & (7,12,1,2)\leadsto^{\trans} 14\\
(15) & c\at \neg p & (14,13)\leadsto^\Box 15
\end{array}
\end{array}
\]
}
\caption{A closed 
tableau  for  $\{\Diamond_{\s}\Diamond_{\s} p
\wedge \Box_{\s}\neg p,~
\trans(\r),~\r\subrole \s,~\s\subrole \r\}$}
\label{ese1}
\end{figure}

\end{example}

\begin{example}
\label{esempio2}
\input esempio2

\end{example}

\begin{example}
\label{siblings}
\input ese-siblings

\end{example}

The correspondence between modal and description logics
\cite{DLcorrespondence} makes the result of this work of interest also
in the context of 
description logics.  In  \cite{conjunctivequery,DescriBinder} 
it is shown that the hybrid binder can play a
 useful role in query answering, considering that its occurrences can be
 restricted so as to guarantee decidability 
\cite{stepmother}. 
 In the cited works, however, the restrictions on the interplay between
the binder and universal quantification is orthogonal to the one
considered in the present work.  Occurrences of the universal
quantifier in the scope of the binder are in fact restricted so that
their scope is, in turn, a negated variable. In modal terms, 
the scope of  a $\Box_{\R}$ 
occurring
 in the scope
of a binder is a negated variable. A formula whose NNF has a
subformula of the form $\binder x. F(\Box_{\R}\neg x)$ will be said to
contain the pattern $\binder\Box\neg x$.

The procedure defined in this work is provably complete and
terminating only when the input formula does not contain the pattern
$\Box\binder\Box$ (see Example 8 in \cite{jar2012} for a case where tableau
construction does not terminate).
The termination proof given in this work cannot easily be extended to
cover occurrences of the pattern $\binder\Box\neg x$, and 
whether the restriction to formulae without the pattern
$\Box\binder\Box$ 
can be relaxed by allowing patterns $\Box\binder\Box\neg x$ is an open
question. Next examples show  however cases where tableau
construction terminates and gives the correct result in
 two simple query answering  given in
\cite{conjunctivequery} (reformulated in modal terms).

\begin{example}
\label{esempio3}
\input esempio3
\end{example}

\begin{example}
\label{esempio4}
\input esempio4
\end{example}

%% file: esempio2.tex

Next example illustrates the dynamic nature of blockings. Figure \ref{example2}
represents a complete and open tableau branch $\ramo$ for the
assertion 
$\trans(\r)$ and
the formula
\[F=\Diamond_{\r}\top   \wedge 
\A \Box_{\r^- }p   \wedge  
\Box_{\r}G\mbox{~~~where~~} G=\binder  x.\Diamond_{\r}\binder  y.x \at
\Diamond_{\r}\neg\,y\] 
When $\r$ is transitive, $F$ holds at a state $w$ of an interpretation
$\M$ if $w$ has at least one $\r$-successor, all its $\r$-descendants
have at least two different $\r$-successors and every state of the
model with at lest one $\r$-successor satisfies $p$.

\begin{figure}[htb]
{
\[
\hspace*{-1em}\begin{array}{lll}
\begin{array}{rlc}
0) &a_0 \at  F\\ 
1) &\trans(\r)\\ 
2) &\r\subrole \r\\ 
3) &a_0 \at  (\Diamond_{\r}\top   \wedge 
     \A \Box_{\r^- }p   )  & 0 \leadsto^\wedge 3  \\ 
4) &a_0 \at  \Box_{\r}G& 0 \leadsto^\wedge 4  \\ 
5) &a_0 \at  \Diamond_{\r}\top   & 3 \leadsto^\wedge 5  \\ 
6) &a_0 \at  \A \Box_{\r^- }p    & 3 \leadsto^\wedge 6  \\ 
7) &a_0 \at  \Diamond_{\r}a_1   & 5 \leadsto^\Diamond 7  \\ 
8) &a_1 \at  \top  & 5 \leadsto^\Diamond 8  \\ 
9) &a_1 \at  \Box_{\r^- }p   & \indietro(6,7) \leadstoA 9  \\ 
10) &a_0 \at  \Box_{\r^- }p   & \indietro(6,0) \leadstoA 10  \\ 
11) &a_1 \at  \Box_{\r}G& \indietro\indietro(4,7,1,2) \leadsto^{\mathsf { Trans }} 11  \\ 
12) &a_1 \at  G& \indietro(4,7) \leadsto^\Box 12  \\ 
13) &a_0 \at  p  & \indietro(9,7) \leadsto^\Box 13  \\ 
14) &a_1 \at  \Diamond_{\r}\binder  y.a_1 \at  \Diamond_{\r}\neg\, y      
            & 12 \leadstobinder 14  \\ 
15) &a_1 \at  \Diamond_{\r}a_2   & 14 \leadsto^\Diamond 15  \\ 
16) &a_2 \at  \binder  y.a_1 \at  \Diamond_{\r}\neg\, y     
          & 14 \leadsto^\Diamond 16  \\ 
17) &a_2 \at  \Box_{\r}G& \indietro\indietro(11,15,1,2) \leadsto^{\mathsf { Trans }} 17  \\ 
18) &a_2 \at  G& \indietro(11,15) \leadsto^\Box 18  \\ 
19) &a_2 \at  a_1 \at  \Diamond_{\r}\neg\, a_2    & 16 \leadstobinder 19  \\ 
20) &a_2 \at  \Diamond_{\r}\binder  y.a_2 \at  \Diamond_{\r}\neg\, y 
     & 18 \leadstobinder 20 \\
21) &a_1 \at  \Diamond_{\r}\neg\, a_2    & 19 \leadsto^@ 21  \\ 
\end{array}
&~~~~&
\begin{array}{rll}
22) &a_2 \at  \Diamond_{\r}a_3   & 20 \leadsto^\Diamond 22 \\
23) &a_3 \at  \binder  y.a_2 \at  \Diamond_{\r}\neg\, y     
           & 20 \leadsto^\Diamond 23  \\ 
24) &a_1 \at  \Diamond_{\r}a_4   & 21 \leadsto^\Diamond 24  \\ 
25) &a_4 \at  \neg\, a_2   & 21 \leadsto^\Diamond 25  \\ 
26) &a_3 \at  \Box_{\r}G     
             & \indietro\indietro(17,22,1,2) \leadsto^{\mathsf { Trans }} 26  \\ 
27) &a_3 \at  G& \indietro(17,22) \leadsto^\Box 27  \\ 
28) &a_3 \at  a_2 \at  \Diamond_{\r}\neg\, a_3    & 23 \leadstobinder 28 \\ 
29) &a_4 \at  \Box_{\r}G& \indietro\indietro(11,24,1,2) \leadsto^{\mathsf { Trans }} 29  \\ 
30) &a_4 \at  G& \indietro(11,24) \leadsto^\Box 30  \\ 
31) &a_2 \at  \Diamond_{\r}\neg\, a_3    & 28 \leadsto^@ 31  \\ 
32) &a_4 \at  \Diamond_{\r}\binder  y.a_4 \at  \Diamond_{\r}\neg\, y      
            & 30 \leadstobinder 32 \\
33) &a_2 \at  \Diamond_{\r}a_5   & 31 \leadsto^\Diamond 33\\ 
34) &a_5 \at  \neg\, a_3   & 31 \leadsto^\Diamond 34  \\ 
35) &a_4 \at  \Box_{\r^- }p   & \indietro(6,24) \leadstoA 35  \\ 
36) &a_3 \at  \Box_{\r^- }p   & \indietro(6,22) \leadstoA 36  \\ 
37) &a_2 \at  \Box_{\r^- }p   & \indietro(6,15) \leadstoA 37  \\ 
38) &a_1 \at  p  & \indietro(35,24) \leadsto^\Box 38  \\ 
39) & a_2\at p  & \indietro(36,22) \leadsto^\Box 39 \\
40) &a_4 \at  \Diamond_{\r}a_6   & 32 \leadsto^\Diamond 40 \\ 
41) &a_6 \at  \binder  y.a_4 \at  \Diamond_{\r}\neg\, y    
                & 32 \leadsto^\Diamond 41  \\ 
42) &a_6 \at  \Box_{\r^- }p   & \indietro(6,40) \leadstoA 42  \\ 
43) &a_4 \at  p  & \indietro(42,40) \leadsto^\Box 43  
\end{array}
\end{array}\]
}
\caption{A complete tableau branch for $\{
\Diamond_{\r}\top   \wedge 
\A \Box_{\r^- }p   \wedge  
\Box_{\r}G,  \trans(r)\}$, where $G=\binder  x.\Diamond_{\r}\binder  y.x \at
\Diamond_{\r}\neg\,y$.}
\label{example2}
\end{figure}

In the comments below,  the notation ${\cal B}_n$ is
used to denote the branch segment up to node $n$ included.
\butta{
 and 
$a_i\approx_na_j$ means that $a_i$ and $a_j$ are compatible in
 $\ramo_n$}
Note that, in this example, the formulae to be taken into account to
check compatibilities are $p$, $\Box_{\r^-}p$ and  $\Box_{\r} G$.

The root nodes are (beyond nodes labelled by assertions): 0--6 and 10,
and the offspring relation is:
\[\begin{array}{lll}
 5\prec_{\cal B}\{7-9,11-14\}&~~~&
 14\prec_{\cal B}\{15-21,37\}\\ 
 20\prec_{\cal B}\{22,23,26-28,31,36,39\}&&
 21\prec_{\cal B}\{24,25,29,30,32,35,38\}\\ 
31\prec_{\cal B}\{33,34\}&&
32\prec_{\cal B}\{40-43\}
\end{array}\]
For instance, node $7$ is the minor premiss of 
the application of the $\trans$
rule producing $11$, and  the minor premiss of the application
of the $\Box$ rule producing $12$ and $13$;
therefore $7$, $11$, $12$ and $13$ are siblings.
Moreover, $7$ is also the first non-phantom node where $a_1$ occurs
 when the $\A$ rule is applied to produce node $9$ focusing
on $a_1$, therefore $7$ is the minor premiss of the inference, thus
one of $9$'s 
siblings. 

As a further example, 
though node $22$ is a phantom in the final branch, it is
not a phantom in $\ramo_{35}$ (see below). 
The branch $\ramo_{35}$ is expanded by an application of the 
 $\A$ rule 
focusing on $a_3$ and producing node $36$. In this branch, $22$ is the first
non-phantom node where  $a_3$ occurs, so it is the minor premiss of
the $\A$ inference and $22$ and $36$ are siblings (in all branch
segments from $\ramo_{36}$ onwards).

In the whole branch $\ramo={\cal B}_{43}$, the nodes $20$ and $32$ are
blocked by 
$14$,
because $a_1$ is compatible with both $a_2$ and $a_4$: the relevant
formulae such nominals label in the final branch are $p$,
$\Box_{\r^-}p$ and  $\Box_{\r} G$.

The fact that $20$ and $32$ are blocked by $14$ intuitively means that 
 $a_2$ and $a_4$ behave ``like'' $a_1$,
However, 
though $a_2$ and $a_4$ are compatible, the presence of node $25$ does
not allow to identify  the states they denote in a model of this open branch.

Being $20$ and $32$ directly blocked in \ramo, all their descendants 
($22,23,26$--$28\comma 31\comma 33\comma 34,36,39$--$43$)
are
phantom nodes in $\ramo$.

However, node $20$ is blocked by $14$ only in $\ramo_{37}$ (where
$a_1$ and $a_2$ label  $\Box_{\r^-}p$ and  $\Box_{\r} G$) and from
$\ramo_{39}$ onwards, when both $(38)\,a_1\at p$ and $(39)\,a_2\at p$
are added.
In particular, $20$ is not blocked in $\ramo_i$ for $i\leq 36$,
therefore, it is expanded, and its descendants can also be expanded (or
used as minor premisses)
till node $39$ is added to the branch.

Analogously, $32$ is blocked by $14$ in $\ramo_i$ only for $35\leq
i\leq 37$ and $i=43$. 
Therefore, for instance, node $40$ is not a phantom in $\ramo_{42}$,
so that it can be used  as the minor premiss of the
application of the $\Box$ rule producing $43$.
Note also that in $\ramo_{38}$, where $20$ is
not blocked, $a_2$ and $a_4$ are compatible, therefore $20$ blocks
$32$ in this branch segment (though $20$ is not an ancestor of $32$
w.r.t. the offspring relation).

In order for node  $31$ to be blocked by $21$, $a_1$, $a_2$ and $a_3$
must be compatible. But when  $a_1$ and $a_2$ are compatible, node
$20$ is blocked, and in such a case $31$, that is one of $20$'s children, is a
phantom. Therefore $31$ is never directly blocked.

The branch is complete: no further expansion are possible without
violating the restrictions 
on blocked nodes.
In particular, in the whole branch:
\begin{itemize}
\item  the $\A$ rule cannot focus on $a_5$, which only occurs
in phantom nodes. 
\item Though nodes $36$ and $42$, obtained by applications of the $\A$
rule, are phantoms, such a rule cannot focus again on $a_3$ and $a_6$,
which only occur in phantom nodes.
\item Though $26$ and $27$ are phantoms, the $\trans$ and $\Box$   rules cannot
use again $22$ as a minor premiss, since it is a phantom too.
\item  Similarly, the other phantom nodes labelled by relational
formulae cannot be used as minor premisses. For instance,  $40$ 
cannot be used as the minor premiss of an
application of the $\Box$ rule, paired with $29$.
\end{itemize}

%% file: ese-siblings.tex
Figure \ref{ese-siblings} illustrates a closed one-branch tableau for
the set
$\{\A \binder x.F \wedge \A G ,r^-\subrole s\}$ where
$F=\Diamond_{r^- }(p  \wedge \Diamond_{r}(\neg\, x   \wedge 
\Diamond_{r^- }(\neg\, p   \wedge 
\Diamond_{r}x  )  )  ) $ and $G=\Box_{s^- }\Box_{s}\neg\, p $.

The formula
$\binder x.F$ is a modal rewriting of the  concept of sibling  (child of the
same mother and father) given as an example in
\cite{stepmother}, if $r$ is interpreted 
as the $has\_child$ relation on a set of
individuals  and $p$ the
$female$ concept. With this reading, 
 $\A \binder x.F$ states that everybody has a sibling.
If in turn the relation $s$ is read as $has\_parent$, 
$\A G$ is the negation of the query ``is there somebody who is a 
parent of a child having a female parent?''

The branch $\cal B$ shown in Figure \ref{ese-siblings} is closed because of
nodes $12$ and $27$. Some expansion rules are applied even if they are
not necessary to complete the  construction, in order to give a
complete picture of the relations linking the nominals occurring in
the branch (represented by the relational nodes $9$, $11$, $14$, $16$,
$19$,
$21$, $23$ and  $24$).

In  $\cal B$, nodes $1-8$ are root nodes 
and the offspring relation is the following:
$8\prec_{\cal B}\{9-13\}$,
  $13\prec_{\cal B}\{14-18\comma 27\}
$ and 
$18\prec_{\cal B}\{19-26\} $. 

\begin{figure}[htb]
\[
\hspace*{-1em}\begin{array}{lll}
\begin{array}{rlc}
1) & a_{1} \at  (\A \binder  x.F  \wedge \A G     )  &  \\ 
2) & r^-\subrole s\\ 
3) & r\subrole r\\ 
4) & s\subrole s\\ 
5) & a_{1} \at  \A \binder  x.F  & 1 \leadsto^\wedge 5 \\ 
6) & a_{1} \at  \A G      & 1 \leadsto^\wedge 6 \\ 
7) & a_{1} \at  \binder  x.F   & (5,1) \leadstoA 7 \\ 
8) & a_{1} \at  F[a_1/x]
    & 7 \leadstobinder 8 \\ 
9) & a_{2} \at  \Diamond_{r}a_{1}   & 8 \leadsto^\Diamond 9 \\ 
10) & a_{2} \at  (p  \wedge \Diamond_{r}(\neg\, a_{1}   \wedge  \\ 
&~~~\Diamond_{r^- }(\neg\, p   \wedge  \Diamond_{r}a_{1}  )  )  )  
      & 8 \leadsto^\Diamond 10 \\ 
11) & a_{1} \at  \Diamond_{s}a_{2}   & (9,2) \leadsto^{\mathsf { Link}} 11 \\ 
12) & a_{2} \at  p  & 10 \leadsto^\wedge 12 \\ 
13) & a_{2} \at  \Diamond_{r}(\neg\, a_{1}   \wedge  \\ 
&~~~\Diamond_{r^- }(\neg\, p   \wedge \Diamond_{r}a_{1}  )  )   & 10 \leadsto^\wedge 13 \\ 
\end{array}
&&
\begin{array}{rlc}
14) & a_{2} \at  \Diamond_{r}a_{3}   & 13 \leadsto^\Diamond 14 \\ 
15) & a_{3} \at  (\neg\, a_{1}   \wedge  \\ 
&~~~\Diamond_{r^- }(\neg\, p   \wedge  \Diamond_{r}a_{1}  )  )  
   & 13 \leadsto^\Diamond 15 \\ 
16) & a_{3} \at  \Diamond_{s}a_{2}   & (14,2) \leadsto^{\mathsf { Link}} 16 \\ 
17) & a_{3} \at  \neg\, a_{1}   & 15 \leadsto^\wedge 17 \\ 
18) & a_{3} \at  \Diamond_{r^- }(\neg\, p   
          \wedge \Diamond_{r}a_{1})   
              & 15 \leadsto^\wedge 18 \\ 
19) & a_{4} \at  \Diamond_{r}a_{3}   & 18 \leadsto^\Diamond 19 \\ 
20) & a_{4} \at  (\neg\, p   \wedge \Diamond_{r}a_{1}  )  
        & 18 \leadsto^\Diamond 20 \\ 
21) & a_{3} \at  \Diamond_{s}a_{4}   & (19,2) \leadsto^{\mathsf { Link}} 21 \\ 
22) & a_{4} \at  \neg\, p   & 20 \leadsto^\wedge 22 \\ 
23) & a_{4} \at  \Diamond_{r}a_{1}   & 20 \leadsto^\wedge 23 \\ 
24) & a_{1} \at  \Diamond_{s}a_{4}   & (23,2) \leadsto^{\mathsf { Link}} 24 \\ 
25) & a_{4} \at  \Box_{s^- }\Box_{s}\neg\, p     & (6,19) \leadstoA 25 \\ 
26) & a_{3} \at  \Box_{s}\neg\, p    & (25,21) \leadsto^\Box 26 \\ 
27) & a_{2} \at  \neg\, p   & (26,16) \leadsto^\Box 27 \\ 
\end{array}
\end{array}
\]
\caption{A closed one-branch tableau for 
$\{\A \binder x.F \wedge \A G ,r^-\subrole s\}$ where
$F=\Diamond_{r^- }(p  \wedge \Diamond_{r}(\neg\, x   \wedge 
\Diamond_{r^- }(\neg\, p   \wedge 
\Diamond_{r}x  )  )  ) $ and $G=\Box_{s^- }\Box_{s}\neg\, p $.}
\label{ese-siblings}
\end{figure}

\butta{

\[\begin{array}{cc}
 &8\prec_{\cal B}\{9, 10, 11, 12, 13\}\\ 
 &13\prec_{\cal B}\{14, 15, 16, 17, 18, 27\}\\ 
 &18\prec_{\cal B}\{19, 20, 21, 22, 23, 24, 25, 26\}\\ 
\end{array}
\]
}

%% file: esempio3.tex
Consider a knowledge base $K$ where a given state has an $\r$-successor
with an $\s$-successor, where $\s$ is transitive and symmetric:
\[K = \{ \Diamond_{r}\Diamond_{s}\top, s^-\subrole s, \trans(s)\}\]
 and
the query ``is there a state which is $\s$-related to itself?'', that
obviously holds in any model of $KB$.
The query is represented by the formula $Q = \E \binder
x.\Diamond_{s} x$, and it is implied from the knowledge base iff 
$K\cup \neg Q$ is unsatisfiable. The NNF of $\neg Q$ ($ \A \binder
x.\Box_{s}\neg\, x$) contains the
pattern $\Box\binder\Box\neg x$.

Let 
$F=\Diamond_{r}\Diamond_{s}\top    \wedge \A \binder
x.\Box_{s}\neg\, x$.  Figure \ref{esempio3-fig} shows a closed one-branch
tableau for $\{F, s^-\subrole s, \trans(s)\}$.

\begin{figure}[htb]
\[
\begin{array}{lll}
\begin{array}{rlc}
0)  & a_0 \at  F\\
1)  & s^-\subrole s\\ 
2) &\trans(s)\\ 
3)  & s\subrole s\\ 
4)  & r\subrole r\\ 
5)  & a_0 \at  \Diamond_{r}\Diamond_{s}\top    &  0 \leadsto^\wedge 5  \\ 
6)  & a_0 \at  \A \binder  x.\Box_{s}\neg\, x      &  0 \leadsto^\wedge 6  \\ 
7)  & a_0 \at  \Diamond_{r}a_1   &  5 \leadsto^\Diamond 7  \\ 
8)  & a_1 \at  \Diamond_{s}\top   &  5 \leadsto^\Diamond 8  \\ 
\end{array}
&~~&
\begin{array}{rlc}
9)  & a_1 \at  \Diamond_{s}a_2   &  8 \leadsto^\Diamond 9  \\ 
10)  & a_2 \at  \top  &  8 \leadsto^\Diamond 10  \\ 
11)  & a_2 \at  \Diamond_{s}a_1   &  (9,1) \leadsto^{\mathsf { Link}} 11  \\ 
12)  & a_2 \at  \binder  x.\Box_{s}\neg\, x     &  (6,9) \leadstoA 12  \\ 
13)  & a_2 \at  \Box_{s}\neg\, a_2    
   &  12 \leadstobinder 13  \\ 
14)  & a_1 \at  \Box_{s^- }\neg\, a_2    
   &  (13,9,2,1) \leadsto^{\mathsf { Trans }} 14  \\ 
15)  & a_2 \at  \Box_{s^- }\neg\, a_2    
  &  (14,11,2,3) \leadsto^{\mathsf { Trans }} 15  \\ 
16)  & a_1 \at  \Box_{s}\neg\, a_2    
  &  (15,11,2,1) \leadsto^{\mathsf { Trans }} 16  \\ 
17)  & a_2 \at  \neg\, a_2   &  (16,9) \leadsto^\Box 17  \\ 
\end{array}
\end{array}
\]
\caption{A closed tableau for 
$\{\Diamond_{r}\Diamond_{s}\top    \wedge \A \binder
x.\Box_{s}\neg\, x, s^-\subrole s, \trans(s)\}$
}
\label{esempio3-fig}
\end{figure}

\butta{
The root nodes are (beyond nodes labelled by assertions): 0, 5, 6,  
and the offspring relation is:

\begin{align*}
 &5\prec_{\cal B}\{7,8\}\\ 
 &8\prec_{\cal B}\{9,10,11,12,13,14,15,16,17\}\\ 
\end{align*}
}


%% file: esempio4.tex
Let $K$ be a knowledge base containing the axiom
$\A (\Diamond_{r}\top   \vee  \Diamond_{s}\top  ) $ where $r$ is a
transitive and symmetric relation:
\[K=\{\A (\Diamond_{r}\top   \vee  \Diamond_{s}\top  ),
r^-\subrole r, \trans(r)\}\]
 and let $Q$ be the query 
$ \E \binder  x.(\Diamond_{r} x \wedge  \Diamond_{s}\Diamond_{r}\binder
z.\Diamond_{r} z)$ (``is there a state that is $\r$-related to itself
and $\s$-related to a state which is in turn $\r$-related to
itself?''), that is not derivable from $K$.

Figure \ref{esempio4-fig} shows a complete and open branch for $K\cup\neg
Q = \{F, r^-\subrole r, \trans(r)\}$, where
$F=\A (\Diamond_{r}\top   \vee  \Diamond_{s}\top  )   \wedge 
\A \binder  x.(\Box_{r}\neg\, x    \vee  G )$, and
$G= \Box_{s}\Box_{r}\binder z.\Box_{r}\neg\, z$.  In that branch, node
20 is blocked by 15, because $a_1$ and $a_2$ are compatible in the
branch (the only relevant nodes to establish it are 24 and 26).
Therefore, every applicable rule has been applied in the branch.

\begin{figure}[htb]
\[
\begin{array}{lll}
\begin{array}{rlc}
0)  &a_0 \at  F\\
1) & r^-\subrole r\\ 
2) & \trans(r)\\ 
3) & r\subrole r\\ 
4) & s\subrole s\\ 
5) & a_0 \at  \A (\Diamond_{r}\top   \vee  \Diamond_{s}\top  )   
  &  0 \leadsto^\wedge 5 \\ 
6) & a_0 \at  \A \binder  x.(\Box_{r}\neg\, x    \vee G) &  0 \leadsto^\wedge 6 \\ 
7) & a_0 \at  \binder  x.(\Box_{r}\neg\, x    \vee G)  &  (6,0) \leadstoA 7 \\ 
8) & a_0 \at  (\Diamond_{r}\top   \vee  \Diamond_{s}\top  )  
   &  (5,0) \leadstoA 8 \\ 
9) & a_0 \at  (\Box_{r}\neg\, a_0    
  \vee  G)
  &  7 \leadstobinder 9 \\ 
10) & a_0 \at  \Box_{r}\neg\, a_0    &  9 \leadsto^\vee 10 \\ 
11) & a_0 \at  \Diamond_{s}\top   &  8 \leadsto^\vee 11 \\ 
12) & a_0 \at  \Diamond_{s}a_1   &  11 \leadsto^\Diamond 12 \\
\end{array}
&~~&
\begin{array}{rlc}
13) & a_1 \at  \top  &  11 \leadsto^\Diamond 13 \\ 
14) & a_1 \at  (\Diamond_{r}\top   \vee  \Diamond_{s}\top  )  
    &  (5,12) \leadstoA 14 \\ 
15) & a_1 \at  \Diamond_{r}\top   &  14 \leadsto^\vee 15 \\ 
16) & a_1 \at  \Diamond_{r}a_2   &  15 \leadsto^\Diamond 16 \\ 
17) & a_2 \at  \top  &  15 \leadsto^\Diamond 17 \\ 
18) & a_2 \at  \Diamond_{r}a_1   &  (16,1) \leadsto^{\mathsf { Link}} 18 \\ 
19) & a_2 \at  (\Diamond_{r}\top   \vee  \Diamond_{s}\top  )  &  (5,16) \leadstoA 19 \\ 
20) & a_2 \at  \Diamond_{r}\top   &  19 \leadsto^\vee 20 \\
21) & a_2 \at  \binder  x.(\Box_{r}\neg\, x    \vee G)
  &  (6,16) \leadstoA 21 \\ 
22) & a_1 \at  \binder  x.(\Box_{r}\neg\, x    \vee G)
  &  (6,12) \leadstoA 22 \\ 
23) & a_1 \at  (\Box_{r}\neg\, a_1    \vee  G)
   &  22 \leadstobinder 23 \\ 
24) & a_1 \at  \Box_{s}\Box_{r}\binder  z.\Box_{r}\neg\, z       
  &  23 \leadsto^\vee 24 \\ 
25) & a_2 \at  (\Box_{r}\neg\, a_2    \vee  G )  
  &  21 \leadstobinder 25 \\ 
26) & a_2 \at  \Box_{s}\Box_{r}\binder  z.\Box_{r}\neg\, z       
&  25 \leadsto^\vee 26 \\ 
\end{array}
\end{array}
\]
\caption{A complete and open branch in a tableau for 
$ \{\A (\Diamond_{r}\top   \vee  \Diamond_{s}\top  )   \wedge 
\A \binder  x.(\Box_{r}\neg\, x    \vee  \Box_{s}\Box_{r}\binder
z.\Box_{r}\neg\, z ), r^-\subrole r,
\trans(r)\}$}
\label{esempio4-fig}
\end{figure}


\butta{
The root nodes are (beyond nodes labelled by assertions): 0, 5, 6, 7,
8, 9, 10, 11, 
and the offspring relation is:

\begin{align*}
 &11\prec_{\cal B}\{12,13,14,15,22,23,24\}\\ 
 &15\prec_{\cal B}\{16,17,18,19,20,21,25,26\}\\ 
\end{align*}
}


%% file: nobinder.tex
\section{In the absence of the binder}
\label{nobinder}

\margin{modif The calculus presented in this work is the first
terminating one dealing with restricted occurrences of the binder."
Maybe you should add that it is though not the first work to present
such a calculus, since [8,9] does the same.}

The calculus presented in Section \ref{calcolo} 
 is the first terminating one dealing with
restricted occurrences  of the binder.
In order to compare it with other works in the literature, its binder
free subsystem has to be considered. In the absence of the binder, the
strong subformula property holds for any node label: if $(n)\,a\at F$
is a node in a tableau branch, then $F$ is 
obtained from a subformula of the top formula $F_0$ of the branch by
 by possibly replacing  
operators $\Box_\R$
with $\Box_\S$, for some relation $\S$ in the language of the initial tableau.
In particular, then, for every node label $a\at F$, $F$ does not
contain any non-top nominal. As a consequence,
if a node $(n)\,a\at F$ blocks $(m)\,b\at F'$, then $F=F'$ and 
$a$ and $b$ label the same
set of propositions in $\PROP$ and formulae of the form $\Box_{\R}G$.

\butta{
Differently from other terminating tableau calculi for (binder-free)
hybrid logic including the global and converse modalities, blocking
concerns here nodes (corresponding to formulae) and not nominals
({\ie} sets of formulae). 
Moreover, as argued above, in the absence of the binder, compatibility
checks, requiring to exit from the ``local'' view and look for other
formulae in the branch, are needed only for the formulae outermost
nominals and concern only a subset of the formulae labelled by such
nominals.  
On the contrary, 
}

Most approaches to blocking in both hybrid logic and
description logic tableaux consist in blocking nominals (or
individuals), taking into consideration the whole
set of formulae they make true in the branch ({\em equality blocking}).
An exception is represented by {\em pattern based blocking} in 
\cite{KaSmoJoLLI2009}, where a terminating system for binder-free hybrid
logic with the global, converse and difference modalities, as well as
reflexive and transitive relations, is defined. Pattern based blocking
blocks formulae ({\ie} nodes, in the setting of the present work),
considering only a subset of the formulae labelled by the involved
nominals (though a larger subset than the one needed to check
compatibilities). 
Pattern based
blocking, however, is applied only in the subcalculus without converse
modalities and termination is not guaranteed 
unless applications of the $\Box$ rule are
prioritized.

The formulation of the $\trans$ rule of Table \ref{regole} 
is very close to the
corresponding one used in description logics, where in fact ``roles''
include both {\em role names} (corresponding to relation symbols) and
the inverse of role names, and inverse roles may also occur in role
inclusion axioms.  The abbreviation $a\Rightarrow_{\R}b$, however,
does not have exactly the same meaning as the corresponding premiss
used in the rule treating transitivity in description logics
\cite{HoSa99,HoSa07a} (a similar approach is adopted in
\cite{KS-graded-role-hierarchies}), consisting of the meta-notion
``$b$ is an $\R$-neighbour of $a$''.
There are two main differences between the two approaches.  First of
all, the semantical notion of accessibility between two states is here
given a ``canonical representation'' in the object language (a choice
already made in \cite{frammento-tab2011,jar2012}): the fact that a
state $a$ is $\r$-related to $b$ is represented by the {\em relational
  formula} $a\at\Diamond_{\r} b$.  Though semantically equivalent to
$b\at\Diamond_{\r^-} a$, the latter is not a relational formula, {\ie} it
is not the canonical representation of an $\r$-relation. This is
reflected by the fact that the $\Diamond$ rule cannot be applied to a
relational formula, while $b\at\Diamond_{\r^-} a$ can be expanded,
producing a relational node. Moreover, in the present work, the
notation $a\Rightarrow_{\R}b$ is only an abbreviation for a relational
formula, which does not take subrelations into account: it may be the
case that $a\Rightarrow_{\S}b$ belongs to a given branch $\ramo$ for
some ${\S}\subrole{\R}$, and yet $a\Rightarrow_{\R}b$ does not.  The
fact that, in the present work, no meta-notion is used to represent
``$\R$-neighbours'' is responsible for the presence of the $\link$
rules, that have no counterpart in
\cite{HoSa99,HoSa07a,KS-graded-role-hierarchies}.

An approach that, in the above respect,
 shares some similarities to the present one is
represented by
\cite{shoi-schmidt}, where a tableau calculus for $\cal SHOI$ is
proposed. In that work, relations between individuals are explicitly
represented
 by use of expressions similar to relational formulae, and,
in fact, the description logic counterpart of the $\link$ rule is
included in the calculus. The calculus however enjoys only a form of
weak termination.

%% file: concl.tex
\section{Concluding Remarks}
\label{concl}

This work presents a satisfiability decision procedure for hybrid
formulae in $\thframmento$\siball{, that has been implemented in the {\sibyl}
prover}. It is also proved that,  
although a restricted use of graded modalities can be
added to the considered fragment whithout endangering decidability, 
in general, their addition to 
 $\Hl_m(@\comma \binder)\setminus \Box\binder
  \Box$ or $\Hl_m(\binder\comma\Diamond^-)\setminus\allowbreak \Box\binder
  \Box$ (in the simple form of functional restrictions) 
 results in logics with an undecidable satisfiability problem.

\sibsmall{The proof procedure
 has been implemented in a
prover called {\sibyl}, that is available at
\url{http://cialdea.dia.uniroma3.it/sibyl/}.  It is written in
Objective Caml and runs under the Linux
operating system.
{\sibyl}
takes as input a file containing a set of assertions and a set
of formulae, checks them for satisfiability and outputs the result.
Optionally, a {\LaTeX} file
with the explored tableau branches can be produced.
Every input formula in $\mmframmento$ is
preprocessed and translated into the fragment $\Hl_m(@\comma
\binder\comma \E\comma \Diamond^-) \setminus\binder\Box$, by use of 
 the 
satisfiability preserving translation defined in Section
\ref{preprocess}.
Some first experiments with the prover were carried out in order to
test it for correctness.  
The test sets, the detailed experimental results 
 and diagrams summarizing them,
as well as some tools used for the experiments,
are available at
 {\sibyl} web page.
A brief summary of the results of the experiments is also reported in
\cite{cade2013}.
}

The core of the proof procedure is a tableau calculus where
transitivity and relation inclusion assertions are treated by
expansion rules which are very close to (though not exactly the same
as) the analogous rules presented in
\cite{HoSa99,HoSa07a,KS-graded-role-hierarchies,KaSmoJoLLI2009}.  The
main result of this work is proving that they can be added to a
calculus dealing also with restricted occurrences of the binder,
maintaining termination, beyond soundness and completeness.

Differently from other terminating tableau calculi for (binder-free)
hybrid logic including the global and converse modalities, blocking
concerns here nodes (corresponding to formulae) and not nominals
({\ie} sets of formulae). In the absence of the binder, compatibility
checks, requiring to exit from the ``local'' view and look for other
formulae in the branch, are needed only for the formulae outermost
nominals and concern only a subset of the formulae labelled by such
nominals.  Indirect blocking, in turn, relies on a particular partial
order on nodes, arranging them in a family of trees of bounded width
and bounded depth.  Width boundedness is guaranteed by the fact that
universal nodes (which may be expanded a potentially unbounded number
of times) do not generate ``siblings''.

Other works have addressed the issue of representing frame properties
and/or relation hierarchies in tableau calculi for binder-free hybrid
logic (for instance,
\cite{BolBlack2,KS-graded-role-hierarchies,KaSmoJoLLI2009}).  The
maybe richer calculus of this kind is
\cite{KS-graded-role-hierarchies}, that considers graded and global
modalities, reflexivity, transitivity and role hierarchies.  The
converse modalities are however missing, and inverse relations are not
allowed.

The scope and interest of the logic considered in this work is widened
by the fact that it subsumes the expressive
description logic $\cal SHOI$.
The possibility of adding limited uses of the binder to description
logics has been addressed, for instance, in
\cite{conjunctivequery,DescriBinder,stepmother}. 
In the cited works, however, the restrictions on the interplay between
the binder and universal quantification is orthogonal to the one
considered in the present work.  Occurrences of the universal
quantifier in the scope of the binder are in fact restricted so that
their scope is, in turn, a negated variable.
The termination proof given in this work cannot easily be extended to
cover occurrences of such a pattern and whether the restriction to
formulae without the pattern 
$\Box\binder\Box$ can be relaxed is an open question.

\butta{
 (in modal terms, 
the scope of  a $\Box_{\R}$ 
occurring
 in the scope
of a binder is a negated variable). 
The termination proof given in this work cannot easily be extended to
cover occurrences of the pattern $\binder\Box\neg x$, and 
whether the restriction to formulae without the pattern
$\Box\binder\Box$ 
can be relaxed by allowing patterns $\Box\binder\Box\neg x$ is an open
question. 
}

\bigskip\noindent
{\bf Acknowledgements.} The author thanks Serenella Cerrito for her
useful suggestions, which helped to give a better presentation of this works.

%% file: appendix.tex

\section{Termination and completeness with transitive relations and
  relation hierarchies}
\label{proofs}

The calculus presented in Section \ref{calcolo}
is  trivially sound. Moreover,
it 
is complete and terminating, {\em provided that the initial formula is in
the fragment 
$\Hl_m(@\comma \binder\comma \E\comma \Diamond^-)
  \setminus\binder\Box$.}
The whole  termination and 
completeness
proofs are quite long already for the calculus defined in
\cite{jar2012},  so they are just summarized in this appendix,
focusing on the
integrations and modifications needed to add  assertions are shown.
In order to make the presentation as readable as possible, however,
statements and definitions are fully reported, when needed to
understand the  changes {\wrt} the proofs given in \cite{jar2012}.  
The numbering of lemmas will be the same as in \cite{jar2012}, so that
the reader can easily find them for comparison, and new intermediate
results and definitions 
are numbered autonomously.

In what follows, it is always assumed that the initial formula of the
tableau is 
in the fragment $\Hl_m(@\comma \binder \comma \E\comma \Diamond^-)
  \setminus\binder\Box$, even when it is not
explicitly stated.

The key result used to prove both termination and completeness is a
form of subformula property.
In the presence of subroles, the definition of the set of subformulae
of a given formula $F$ has to be widened, allowing for relation
renaming (because of the $\trans$ rule), {\ie} the subformulae of $F$ 
include 
 all the formulae of the form 
 $\Box_\R G$ 
for every subformula 
$\Box_\S G$ 
of $F$ and for every 
relation  $\R$ in the language.

\begin{defi}\label{sottoformule}
If $F$ is a hybrid formula and $\REL$ a set of relation symbols,
 then $G$ is a subformula of $F$ {\wrt} $\REL$ if and only if
either $F=G$ or one of the following conditions holds:
\begin{itemize}
\item $F=F_1\star F_2$, for $\star\in\{\wedge,\vee\}$ and $G$ is a
  subformula of $F_i$;
\item either $F=t\at F_0$ or $F=\binder x.F_0$ or $F=\nabla F_0$ for
  $\nabla\in\{\A,\E,\Diamond_\R\}$, and $G$ is a
  subformula of $F_0$;
\item $F=\Box_{\R} F_0$, for some {\em relation} $\R$, 
and $G$ is a  either a subformula of $F_0$ or
$G=\Box_{\S} F_0$  for some {\em relation} $\S$, for $\s\in\REL$.
\end{itemize}

If $\ramo$ is a tableau branch and $a_0\at F_0$ its top formula, 
$\Subf(\ramo)$ is the set of the subformulae of $F_0$ {\wrt} the set
$\REL$ 
of relation symbols occurring in 
 the initial tableau and
\[
\Cmp(\ramo)= 
(\Subf(\ramo) \cap \PROP)\cup
\{\Box_{\R} G\mid \Box_{\R} G\in \Subf(\ramo)\}
\]
\end{defi}

The following result bounds the number of subformulae of a given
formula.

\begin{newlemma}\label{contasub}
Let $F$ be a formula in a language with $M$ relation symbols (which do
not necessarily occur all in $F$),
and $\left|F\right|=N$ the size of $F$.
Then $F$ has  no more than $2\times M\times N$ subformulae.
\end{newlemma}

\begin{proof}
The number of subexpressions of $F$ is bounded by $N$.
From each subexpression of the form form $\Box_\R G$, 
$2\times M$
more subformulae of $F$ can be obtained.
 Therefore, $F$ has no more than $2\times M\times N$
subformulae.
\end{proof}

\butta{
Serve anche che  se $\Diamond_R b$ \`e
una sottoformula di $G$ allora anche $\Diamond_S b$ \`e
una sottoformula di $G$ (dove $b$ \`e un nominale)? Mi sembra di no,
nel lemma sono escluse le relazionali.
}
With this modification, the main property of the calculus, 
the (weak and strong) subformula property, still holds.
It uses the notion of {\em instance} of a formula $F$, that is 
 any expression obtained by uniformly replacing every free
variable in $F$ with some nominal. 

\setcounter{lemma}{3}

\begin{lemma}[Subformula properties]\label{subformula}
For any formula $a\at F$ occurring in a branch ${\cal B}$ which is not
a relational formula, $F$ is an instance of a formula in
$\Subf(\ramo)$ (weak subformula property).

Moreover, assuming that the initial formula of the branch is 
in the fragment with no patterns $\binder\Box$,
 if $F$ is a universal formula, then $F\in \Subf(\ramo)$ (strong
 subformula property). 
\end{lemma}

\begin{proof}
The proof is an induction on the construction of $\ramo$, which
simultaneously proves the following strongest versions of the two
properties: if $(n)\,a\at F$ is a node in ${\cal B}$ and $a\at F$ is
not a relational formula, then for any subformula $F'$ of $F$:
\begin{itemize}
\item[ ({$\alpha$})]  $F'$ is an instance of a formula  in $\Subf(\ramo)$, and 
\item[({$\beta$})] 
if $F'$ is a universal formula,
then $F'\in \Subf(\ramo)$.
\end{itemize}
 
The induction step of the corresponding proof in \cite{jar2012} can
  easily be extended  with the cases where the
branch $\ramo$ is obtained from $\ramo'$ by
application of one of the new rules. We show below the treatment of
the the $\link$ and $\trans$
rules (the extension to the multi-modal case of the other rules is
straightforward).

\begin{enumerate}

\item If $\ramo$ is obtained by application of the $\link$ rule, then
  there is nothing to prove since
  the newly added node is labelled by a relational
  formula.

\item If $\ramo$ is obtained by application of the $\trans$
rule, then {$\alpha$} and {$\beta$}
directly follow from the induction hypothesis.
\qedhere
\end{enumerate}
\end{proof}

%% file: termina.tex
\subsection{Termination}
\label{termina}

Termination of the calculus presented in Section \ref{calcolo} is proved,
like in \cite{jar2012},
 by showing that the nodes of a branch $\ramo$
are arranged by the offspring relation  into a bounded sized set of
trees, each of which has bounded width and bounded depth.  Hence any
tableau branch $\ramo$ has a number of nodes that is bounded by a
function of the size of the initial tableau.

In order to show  that, in the forest of
trees induced by the offspring relation on the nodes of a branch \ramo,
any node has a bounded number of siblings, the key result  
 is Lemma \ref{base} below. 
It is worth noticing that if a branch were a set of formulae,  this
result would be straightforward. But since the same formula may label
different nodes, things are more involved.

The notation $m\produce n$ denotes the relation
 holding between two nodes $m$ and $n$ whenever they are siblings
 {\wrt} the offspring relation and $n$ has been added to the branch by
 application of an expansion rule to premisses including $m$. {\Ie}
$m\produce n$ if one of the following conditions hold:
\begin{itemize}
\item $n$ is added to the branch by application of one of the rules
  $\wedge,\vee, @,\binder$;
\item $n$ is added to the branch by application of
either a universal rule or the  $\trans$ rule whose minor premiss is $m$;
\item $n$ is added to the branch by application of
a  $\link$ rule whose logical premiss is $m$.
\end{itemize}
The relation 
$\produce^*$ is the reflexive and transitive closure of $\produce$.
If $n\produce^*m$, we say that $n$ {\em produces} $m$.

Below, the notion of subformula is used without explicit reference to
the set $\REL$ of relation symbols that can be used, assuming that 
it is the set of relation symbols occurring in the
initial tableau.

The proof of Lemma \ref{base} uses the notions defined as follows.
Let $M$ be a set of nominals, $F$ a formula (possibly containing free
variables) and $\Delta$ a set of formulae.
\begin{enumerate}
\item $\clo(\Delta)$ (the {\em   closure} of $\Delta$) is the set
  containing all the subformulae 
of every formula in $\Delta$.
\item An $M$-instance of $F$ is a ground formula that can be obtained 
   from  $F$ by replacing its free variables with elements of $M$.
\item The set $\Delta^M$ is the set
  containing all the $M$-instances of every element of $\Delta$.
\end{enumerate}
Note that, though the above definitions are formally the same as in
\cite{jar2012}, the set denoted by $\clo(\Delta)$ is larger, because
of the new notion of subformula. 

If $F$ is a formula and $\calA$ a set of assertions involving
$n$ relation symbols, then  the size of $F$ plus the number of
relation symbols occurring in $\calA$ will be denoted by
$\left| F +\calA\right|= 
\left| F\right| + n$.

\begin{lemma}\label{base}
Let $n$ be a node in a branch $\ramo$ of a tableau for  $\{F\}\cup\calA$,
and let $N=\left| F+\calA\right|$.  Then the cardinality of
$\Sigma(n)=\{m\mid n\produce^*m\}$ is bounded by an exponential
function $E_w(N)$.
\end{lemma}

\begin{proof}
The guiding intuition of the proof consists in  showing that the label of
any node in $\Sigma(n)$ has a {\em matrix} taken from a bounded stock
of formulae, that is built in the language of the branch {\em at the
  time $n$ is added to it}.  Node labels with the same matrix are
always equal, at any construction stage of the branch, so that the
cardinality of $\Sigma(n)$ is bounded by the number of such possible
matrices, since siblings always have the same phantom/non-phantom status.

In order to properly define matrices, some more notations are
introduced below.
Any branch $\ramo$ in a tableau is the last element of a sequence of
branches, where the first one is the initial tableau, and each of the
others is obtained from the previous one by application of an
expansion rule.  Such a sequence will be called the {\em sequence of
branches leading to \ramo}.

Let $n$ be any fixed node in a tableau branch {\ramo}.  
Then:
\begin{enumerate}
\item 
$\ramo_1$ is the first branch where $n$ occurs, in the sequence of
  branches leading to $\ramo$.
\item  $\operatorname{label}_{{\cal {B}}_i}(k)$ is  the label of
the node $k$ in the branch $\ramo_i$.  This allows one 
to refer to node labels in different branches.

\item 
For $1\leq i\leq p$, $\sigma_i$ is the composition of the 
substitutions applied in the sequence $\ramo_1,\dots , \ramo_i$, by
means of the equality rule.  
\butta{
Consequently,
for each $i>0$, $\operatorname{label}_{{\cal B}_i}(n)=
\sigma_i(\operatorname{label}_{{\cal B}_1}(n))$.
}
\item  $M_n$ is the set containing all the nominals occurring in 
$\operatorname{label}_{{\cal B}_{1}}(n)$ and all the top nominals in
  $\ramo_{1}$. 

\item  
$\Gamma_n$, $\Delta_n$ and $S_n$ are the sets of formulae defined as
  follows.
\[\begin{array}{cll}
\Gamma_n &=&\{F\mid F \mbox{ is a universal subformula of the top formula of }
                     \ramo_{1}\}\\
\Delta_n &=&\{\operatorname{label}_{{\cal B}_{1}}(n)\}\cup\Gamma_n
\\
S_n &=& (\clo(\Delta_n))^{M_n}
\end{array}\]
{\ie} $S_n$ contains all the $M_n$-instances of every formula in the
closure of $\Delta_n$.

\butta{
We remark here that, in the new setting of this work, if either
 $\Box_RG$ or $\Box^-_RG$ 
is a subformula of the top formula, then so is also any
formula of the form $\Box_S G$ and $\Box^-_SG$.
}
\item $\F_n$ is the set defined as follows:
\[
\begin{array}{lll}
\F_n&=&\{a\at F\mid a\in M_n\mbox{ and }F\in S_n\}\cup\\
&&
\{a\at \Diamond_\r b\mid a,b\in M_n\mbox{ and }\r\in \REL\}
\end{array}
\]
  Any element of $\F_n$ will be
called a {\em matrix} (this definition is a straightforward extension to
the multi-modal case of the corresponding definition in \cite{jar2012}).
\end{enumerate}

The bound $E_w(N)$ on the cardinality of $\Sigma(n)$, computed in
\cite{jar2012}, is equal to $\left|\F_n\right|$. Such a value, in
turn, is shown to be equal to 
$M+K^2\times N^{N+1}$,
where $M$ 
is the maximal number of relational formulae which can be built out of
$N$ nominals, and $K$ 
is the maximal number of subformulae of a
formula of size $N$.  
In the uni-modal case, $M=N^2$ and $K=N$, 
while in the
multi-modal case $M=N^3$ and $K=2\times N^2$ 
(by Lemma 
\ref{contasub}, since the
number of relation symbols in the language 
is also bounded by $N$). The computation of the exponential factor is
independent of the number of modalities in the language.
Therefore, the bound $E_w(N)$ is exponential also in the
multi-modal case. 


Let $m$ be any node in $\Sigma(n)$, {{\ie}} $n\produce^*m$.  An
element $F$ of $\F_n$ is called a {\em matrix} of $m$ in $\ramo_i$ if
$\operatorname{label}_{{\cal B}_{i}} (m)=\sigma_i(F)$; and $F$ is a
matrix of $m$ if it is a matrix of $m$ in all $\ramo_i$ where $m$
occurs, for $i=1,\dots ,p$.  If two nodes $m_1$ and $m_2$ have the same
matrix, then obviously for all $i=1,\dots ,p$ such that both $m_1$ and
$m_2$ are in $\ramo_i$, $\operatorname{label}_{{\cal B}_{i}}
(m_1)=\operatorname{label}_{{\cal B}_{i}} (m_2)$.

The proof that the cardinality of $\Sigma(n)$ is bounded by
$E_w(N)$, where $E_w(N)$ is the cardinality of $\F_n$, is based on the
fact that every node in $\Sigma(n)$ has a matrix:

\begin{itemize}
\item[($\alpha$)] 
the label of any node in $\Sigma(n)$ has a matrix in $\F_n$. {{\Ie}}
 if $m\in \Sigma(n)$, then there exists  $F\in \F_n$ such that
for all $i\geq 1$,  if $m\in\ramo_i$ then
 $\operatorname{label}_{{\cal B}_{i}}(m)=\sigma_i(F)$.
\end{itemize}

\noindent
The proof is by induction on $i$. We show next the cases of the
induction step corresponding to the $\trans$ and $\link$ rules.
The treatment of the $\Box$ rule is an easy multi-modal rewriting of
the corresponding case in \cite{jar2012} and, as a matter of fact, it
is very similar to the treatment of the $\trans$ rule shwon below.

\begin{description}

\item[($\link$)] Let $n\produce^*k$ and $m$ be obtained by an
  application of the $\link$ rule to nodes $k$ and
  $t$,
  with
\[\begin{array}{l}
\operatorname{label}_{{\cal  B}_{i-1}}(k)=\operatorname{label}_{{\cal
      B}_{i}}(k)=
a\Rightarrow_\R b\\
\operatorname{label}_{{\cal  B}_{i-1}}(m)=\operatorname{label}_{{\cal
    B}_{i}}(m)= a\Rightarrow_{\S} b. 
\end{array}\]
By the induction hypothesis, 
$a\Rightarrow_\R b=\sigma_i(c\Rightarrow_\R d)$ for some $c\Rightarrow_\R d\in
\F_n$, {\ie},
$a=\sigma_i(c)$ and 
 $b=\sigma_i(d)$ for $c,d\in M_n$.
Since $c$ and $d$ belong to $M_n$, 
$c\Rightarrow_{\S} d\in\F_n$. 
Therefore $c\Rightarrow_{\S} d\in\F_n$ 
is a matrix of $a\Rightarrow_{\S} b$  
in $\ramo_i$, because
$a\Rightarrow_{\S} b=\sigma_i(c\Rightarrow_{\S} d)$.

\item[($\trans$)] Let $n\produce^*k$ and $m$ be obtained by an
  application of  the $\trans$ rule to nodes $k,k',t$ and $i$,
  with
\[\begin{array}{l}
\operatorname{label}_{{\cal  B}_{i-1}}(k')=\operatorname{label}_{{\cal
      B}_{i}}(k')=  a\Rightarrow_{\R} b,\\  
\operatorname{label}_{{\cal  B}_{i-1}}(m)=\operatorname{label}_{{\cal
     B}_{i}}(m)=b\at\Box_{\R} G. 
\end{array}\]
 By Lemma \ref{subformula}, 
$\Box_{\R} G\in S_n$. By the induction hypothesis, 
$a\Rightarrow_{\R} b=\sigma_i(c\Rightarrow_{\R} d)$
for some 
$c\Rightarrow_{\R} d\in
\F_n$, {\ie} $b=\sigma_i(d)$ for some $d\in M_n$.  Therefore
$d\at \Box_{\R} G\in \F_n$ and, since
$b\at \Box_{\R} G=\sigma_i(d)\at \Box_{\R} G=\sigma_i(d\at \Box_{\R} G)$, 
$d\at \Box_{\R} G$ is a matrix of $m$ in $\ramo_i$.
\end{description}

The fact that the cardinality of $\Sigma(n)$ is bounded by
$E_w(N)$, where $E_w(N)$ is the cardinality of $\F_n$ is finally
proved like in \cite{jar2012}.
Let us assume,
by {\em reductio ad absurdum}, that $\Sigma(n)$ has more than $E_w(N)$
elements.  Then, by $\alpha$, there are at least two distinct elements
$m_1$ and $m_2$ in $\Sigma(n)$ which have the same matrix $F$.  We may
assume \mbox{\em w.l.g.}  that $n\leq m_1<m_2$.  Let $\ramo_{k}$ be
the first branch  where
$m_2$ occurs.  Since $n<m_2$, there is a node $k\in \Sigma(n)$ such
that $n\,\produce^*k\,\produce m_2$.  Given that $k$ produces a node,
it is not the major premiss of a universal rule or one of the $\trans$
rules.  
Moreover, $k$ is not
a phantom in $\ramo_{{k-1}}$, otherwise  restriction
{\bf R3} would be violated.
  Consequently,  $m_1$ is not a phantom in
$\ramo_{{k-1}}$ either.  But $\operatorname{label}_{{\cal B}_k}(m_2)=
\sigma_k(F)=\sigma_{k-1}(F)=\operatorname{label}_{{\cal
    B}_{k-1}}(m_1)$ ($\sigma_k=\sigma_{k-1}$ because, clearly,
$\ramo_{k-1}$ has not been expanded by means of the equality rule,
which does not add new nodes to the branch).  Therefore, the addition
of $m_2$ to $\ramo_{{k-1}}$ violates restriction {\bf R1}.
\end{proof}

Lemma \ref{base} allows for establishing that the number of trees in the
forest induced by the offspring relation on the nodes of a tableau
branch is bounded by an exponential function of the size of the
initial formula, and so is the
width of each of such trees. Obviously,  the trees include the
single-node ones constituted by nodes labelled by assertions
(assertions do not produce any node), whose number is polynomial
in the number of relations occurring in the initial tableau.

In order to prove that tree depth is also bounded, it is shown that
the size of any set of blockable nodes which may occur in a tableau
branch, and such that none of its elements blocks another one, is
bounded.  This holds essentially because of Lemma \ref{base}:
first of all, the weak subformula property ensures that the number of
possible ``schemata'' for blockable formulae is bounded. 
Secondly, the strong subformula property ensures that 
the number of nominal compatibility classes is also bounded.

In \cite{jar2012}, it is  shown that tree depth is  bounded by
an exponential function $E_d(N)$ of $N$, where $N$ is the size of the
initial formula.  The value of such a function
is computed as $M\times E_d'(N)$, where $M$ is the maximal
cardinality of the set of
subformulae of the initial formula and
$E_d'(N)$ is an exponential function of $N$.
 In the mono-modal case,
$M=N$, while in the multi-modal one, $M=2\times N^2$.
Consequently, both tree width and tree depth increase only of a
polynomial factor {\wrt} the uni-modal case.
 
The rest of the termination proof is independent of 
the presence of 
the new expansion rules and multi-modalities,  therefore,
modulo the replacement of the exact values of
$E_w(N)$ and $E_d(N)$, it  stays the same and the overall
result does not change. Consequently, the termination theorem can be
proved like in \cite{jar2012}.

\begin{theorem}[Termination]
\label{termination}
If the initial formula of a tableau is in the fragment
$\miniframmento$, then every tableau branch 
has a bounded depth and  tableau construction
always terminates.
\end{theorem}



%% file: completeness.tex

\subsection{Completeness}
\label{completeness}

In order to prove that the calculus 
is complete, it is shown -- like in \cite{jar2012} -- how to extend a subset
$\N_0$ of any complete and open branch {\ramo} in such a way that
every directly blocked node is added a suitable ``witness''.  
The witness(es) of a blockable node $n$ can be viewed simply as 
node(s) which could by obtained
by application of the corresponding blockable rule to
$n$.\footnote{Actually, in \cite{jar2012}, witnesses are nominals
  and not nodes, but this detail  can be ignored here.}
The
label of each newly added node is obtained from a node in $\N_0$ by
suitably renaming non-top nominals.  A model of the initial formula
can then be extracted from such an extension. Due to the presence of
assertions,  
 the construction of the model differs
  from the corresponding one in
 \cite{jar2012}. 

The fact that the labels of blocked and blocking nodes are not necessarily
identical does not allow taking
 the witness of the blocking node
as a witness of the blocked one. Nor can a model be simply built from
a set of states consisting 
of equivalence classes of nominals, where two nominals are in the same
class whenever some blocking mapping maps one to the other: two
nominals $a$ and $b$ may be compatible even if  the branch contains a
node labelled by $a\at\neg b$ (this is the case, for instance, of the
nominals $a_2$ and $a_4$ in example \ref{esempio2} of Section
\ref{examples}). 

The set $\N_0$ is the
union of the non-phantom nodes in {\ramo} and the nodes of the form
$(n)\,a\at F$, with $a$ occurring in some
non-phantom node in {\ramo} and $F\in\PROP$ or of 
the form $\Box_{\R} G$.

The extended set $\N_{\cal B}^\infty$ is built by stages, as the union of a
(possibly infinite) sequence of finite extensions of
$\N_0$: $\N_0\subseteq \N_1\subseteq
\N_2\dots$. 
Each set $\N_{i+1}$ is obtained from $\N_i$ by (fairly) choosing a 
blockable 
node $n$ in
$\N_i$ with no witnesses in $\N_i$. The construction ensures that 
there exists a non blocked node $m\in\N_0$ whose label can be mapped to 
$\operatorname{label}(n)$ in $\N_i$.  
Therefore $m$ has been expanded in $\ramo$, generating node(s) with a
fresh nominal $b$. The blocking mapping is then
used to add new nodes and obtain $\N_{i+1}$, in such a way that $n$
has a witness in $\N_{i+1}$: in detail,
let $\pi_i$ be the mapping which maps $m$ to $n$ and $b_i$ a new
nominal. Then a ``nominal renaming'' $\theta_i$ is defined, extending
$\pi_i$ with $b\mapsto b_i$, and  $\N_{i+1}$ extends $\N_i$ by
addition of 
new nodes, each of which is obtained
from a node $k\in\N_0$
by application $\theta_i$ to its label.\footnote{The nominal
  $b$ freshly introduced when expanding $m$ may later on be replaced
  by another nominal $c$, already occurring in the branch when $m$ is
  expanded. In such cases, $b_i$ is not a fresh nominal but
  $b_i=\pi(c)$. This detail can however be ignored here.}
The construction ensures that, for any new node $(k^i)\,F_i$ added at
stage $i$, there exists a node $(n)\,F\in\N_0$ with $F_i=\theta_i(F)$.
Moreover, the extension $\N_i$ is such that it contains a node
labelled by $\theta_i(F)$ for every $F$ occurring as a node label in $\N_0$.
\butta{
If a node is added to $\N_i$ and its label is obtained by
application $\theta_i$ to $k\in\N_0$, it will be denoted by $k^i$.
}

Possibly, new nodes with no witnesses are added, but each of them is
blocked by a (non blocked) node in $\N_0$.  All the ``blocked'' nodes in
$\N_i$ are stored in {\em the blocking relation for $\N_i$}, $\B_i$,
containing triples of the form $(n,m,\pi)$, where $n$ is the
blocked node (a blockable node without witnesses in $\N_i$), 
$m\in\N_0$ is not
blocked, and $\pi$ is a mapping such that 
 $\pi(\operatorname{label}(m))=\operatorname{label}(n)$.

Since the strategy to choose the  nodes to be ``unblocked'' is fair,
the set $\N_{\cal B}^\infty$ is such that every blockable node has
its witness(es).

The construction enjoys the following properties, which can be proved
like in \cite{jar2012} (P\ref{occorre} is stated as Lemma 10 in
\cite{jar2012}, and P\ref{treitems-uno}--P\ref{treitems-tre} constitute 
Lemma 11 in \cite{jar2012}):
\begin{properties}
\item
For all $i$, the renaming $\theta_i$ is an injective function, hence
its inverse $\theta_i^-$ is defined.
\item \label{occorre}
If a nominal $b$ occurs in $\N_0$, then it occurs in some non-phantom
node in $\ramo$. 
\item \label{treitems-uno}
 If $i>0$ and $d$ is a nominal occurring in
  $\N_{i-1}$, then no new node  added at stage $i$ has a label of
  the form $d\at p$ for $p\in\PROP$, or $d\at \Box_{\R} G$.  
As a consequence, if two nominals
  occurring in $\N_{i-1}$ are compatible in $\N_{i-1}$, then, for any $i>0$,
  they stay compatible in $\N_{i}$ (and in $\N^\infty_{\cal B}$).

\item \label{treitems-due} If $i>0$ and $\theta_i$ is the mapping used to
  extend $\N_{i-1}$ to $\N_i$, then for every nominal $d$ occurring in
  $\N_i$, $d$ and $\theta_i(d)$ are compatible in $\N_i$.

\item \label{treitems-tre} For every
triple $(n,m,\pi)\in \B_i$ 
({\ie} the node $n\in\N_i$ is ``blocked'' by $m\in \N_0$ by means of the
mapping $\pi$)
and for every
  nominal $d$ occurring in $\N_i$, $d$ and $\pi(d)$ are compatible in
  $\N_i$.
\end{properties}

In order to build a model of $\N_{\cal B}^\infty$, each of the sets
$\N_i$ is 
shown to enjoy a form of {\em
  saturation property} for non-phantom nodes: it is consistent (there
are no labels of the form $a\at\neg a$, or both $a\at p$ and $a\at\neg
p$), it does not contain non-trivial equalities, and, for any node or
pair of nodes in $\N_i$ that could be the premiss(es) of some
expansion rule other than blockable ones, its expansion(s) are also in
$\N_i$.
Such a notion is defined below.
The definition is the same as in \cite{jar2012}, but for the
reformulation of items  
 \ref{multidiamond} and \ref{multibox} to the multi-modal case and
the addition of the last items
\ref{rel1}--\ref{trans}. 

We abuse notation, writing $F_1,F_2,\dots\in\N_i$, meaning that there
exist
nodes in $\N_i$ labelled by $F_1,F_2,\dots$, respectively

\setcounter{definition}{13}
\begin{definition}
\label{pseudosat}
Let $\ramo$ be a complete and open branch and $\N_i$  an element of
the sequence leading to the construction of $\N_{\cal B}^\infty$. 
The set $\N_i$ is {\em pseudo-saturated} with respect to
$\B_i$ if it satisfies the following properties:

\begin{enumerate}
\item no node in $\N_i$ is labelled by a formula of the form $a\at \neg a$;
\item there are no pairs of nodes labelled by formulae of the form
  $a\at p$ and $a\at \neg p$, for $p\in \PROP$;
\item \label{uguaglianze} if any node in $\N_i$ is labelled by a
  formula of the form $a\at d$ (where $a$ and $d$ are nominals), then
  $a=d$;
\item \label{quattro} if $a\at F\wedge G\in \N_i$ then, 
  $a\at F\in \N_i$ and $a\at G\in \N_i$;
\item if $ a\at F\vee G\in \N_i$, then either
  $a\at F\in \N_i$ or $a\at G\in \N_i$;

\item if $a\at d\at F\in \N_i$, then $d\at
  F\in \N_i$;
\item \label{sette-sat} if $a\at \binder x.F\in \N_i$,  then
  $a\at F[a/x]\in \N_i$;

\item\label{multidiamond}
 if $(n)\,a\at \Diamond_{\R} F\in \N_i$ where
$a\at \Diamond_{\R} F$ is not a relational formula,
 and
 $\B_i$ contains no triple of the form $(n,n',\pi)$
({\ie}  $n$ is not blocked in 
  $\B_i$), then
$a\Rightarrow_{\R}d,~d\at F\in \N_i$,  for some
  nominal $d$
({\ie} $n$ has a witness in $\N_i$);
\item 
\label{multibox}
If $a\at \Box_{\R} F,~
a\Rightarrow_{\R}d\in\N_i$,
then $d\at F\in
  \N_i$.

\item\label{dodici} if $(n)\,a\at \E F\in \N_i$ and $\B_i$ contains no
  triple of the form $(n,n',\pi)$, then 
$d\at F\in \N_i$ for some nominal $d$ ({\ie}  $n$ has a witness in
  $\N_i$);

\item\label{tredici} if $a\at \A F\in \N_i$ and $d$ occurs in $\N_i$,
 then $d\at F\in \N_i$;
\item\label{rel1} $\r\subrole \r\in \N_i$ for all $\r\in\REL$.
\item \label{rel2} If $\R\subrole \S,~\S\subrole \P\in\N_i$, then 
$\R\subrole \P\in\N_i$.

\item \label{link}
if $a\Rightarrow_{\R} b,~R\subrole\S \in\N_i$,
then
$a\Rightarrow_{\S} b\in\N_i$;

\item \label{trans}
if $a\at\Box_{\S}F,~a\Rightarrow_{\R}b,~
\trans(R),~\R\subrole \S\in\N_i$, then
  $b\at\Box_{\R} F\in\N_i$.
\end{enumerate}
\end{definition}


\setcounter{lemma}{11}
\begin{lemma}
\label{saturation_property}
Let $\ramo$ be a complete and open branch and $\N_i$  an element of
the sequence of extensions leading to $\N_{\cal B}^\infty$.
Then $\N_i$ is pseudo-saturated with respect to
$\B_i$.
\end{lemma}
\begin{proof}
First of all, we observe that clauses \ref{rel1} and \ref{rel2} hold
because, since $\ramo$ is complete,  all rules of Table
\ref{assertion-rules} have been applied as far as possible when
building the initial tableau. They generate
root (hence non-phantom) nodes, which 
belong to $\N_0\subseteq \N_i$ for all $i$.

For the other clauses, the proof is by induction on $i$. 
Both the base
case and the induction step of the corresponding proof in
\cite{jar2012} (possibly reformulated for the multi-modal case)
must be completed with the new cases: 
\ref{link} and \ref{trans}, whose treatment is shown below.
Case \ref{link} is quite simple, like 
cases \ref{quattro}--\ref{sette-sat} in \cite{jar2012}, and  
case \ref{trans} is treated very similarly to case \ref{multibox},
 in both the base case and the induction step.

In the induction step, obviously, the pseudo-saturation property in
${\cal N}_{i}$ still holds for all nodes already belonging to 
${\cal N}_{i-1}$. Therefore it must only be shown that the newly added
nodes do not spoil pseudo-saturation.

\begin{enumerate}
\setcounter{enumi}{13}
\item
\begin{description}[font=\normalfont\em]
\item[Base.] If $(n)\,a\at\Diamond_R b\in\N_0$, then
$n$ is not a phantom in $\ramo$. If also $R\subrole\S
  \in\N_0\subseteq\ramo$ 
and $\N_0$ did not contain  $a\Rightarrow_{\S} b$,
then any node labelled by $a\Rightarrow_{\S} b$ in $\ramo$ 
(if present) would be a
phantom.  Therefore, in order for $\ramo$ to be complete, the $\link$
rule should be applied, generating a node $(m)\,a\Rightarrow_{\S} b\in
\ramo$.  
Since $n$ and $m$ would be siblings {\wrt} the offspring
relation, $m$ would not be a phantom in $\ramo$,
therefore $m\in\N_0$.

\item[Induction Step.]
Let $(n^i)\,a\at\Diamond_R b$ be a new node added at stage $i$, and
let $c\at\Diamond_R d$ be the label of the node $n\in\N_0$ such that
$\theta_i(c\at\Diamond_R d)=a\at\Diamond_R b$.
Let us assume that $\R\subrole\S \in\N_i$, hence also
$\R\subrole\S \in\N_0$. 
If $R\subrole\S\in\N_0$, then, since $\N_0$ is pseudo-saturated, it
contains a node labelled by $c\Rightarrow_{\S}d$. Consequently, $\N_i$
contains a node labelled by
$\theta_i(c\Rightarrow_{\S}d)=a\Rightarrow_{\S} b$.
\end{description}
\item 
\begin{description}[font=\normalfont\em]
\item[Base.] If $(m)\,a\Rightarrow_{\R} b\in\N_0$, then 
$m$ is not a phantom in $\ramo$. If also
$a\at\Box_{\S}F\comma ~
\trans(R)\comma ~\R\subrole \S\in\N_0$, then the $\trans$ rule has been
  applied, generating $(k)\,b\at\Box_{\R}F$; $k$ is a sibling of $m$,
  hence non-phantom too, and belongs to $\N_0$.

\item[Induction Step.] Let us assume that 
$\trans(\R)\comma ~\R\subrole \S\in\N_0$, that
  $(n)\,a\at\Box_{\S} F\comma ~(m)\,a\Rightarrow_{\R} d\in\N_i$ and at least one
  of $n$ and $m$ does not belong to $\N_{i-1}$ (otherwise the thesis
  follows from induction hypothesis).
 By Lemma $\ref{subformula}$, $F$ does not
  contain any non-top nominal, hence $\theta_i(F)=F$ for any $i$.

We distinguish two cases: 

\begin{enumerate}
\item $a\at \Box_{\S} F\not\in \N_{i-1}$.  By Property
  P\ref{treitems-uno}, then, $a=b^i$ is the new nominal introduced at stage
  $i$.  Therefore, $\N_0$ contains nodes labelled by
  $\theta^-_i(b^i\at \Box_{\S} F)= \theta^-_i(b^i)\at \Box_{\S} F$ and
  $\theta^-_i(b^i\Rightarrow_{\R} d)=\theta^-_i(b^i)\Rightarrow_{\R}
  \theta_i^-(d)$.  Since $\N_0$ is pseudo-saturated, $\theta_i^-(d)\at
  \Box_{\R}F\in \N_0$, so that $\theta_i(\theta_i^-(d))\at \Box_{\R} F=
  d\at \Box_{\R} F\in \N_i$.
\item $a\at \Box_{\S}F\in \N_{i-1}$.  If  $a\Rightarrow_{\R}
  d\not\in \N_{i-1}$, then  $\theta_i^-(a)\Rightarrow_{\R}
  \theta_i^-(d)\in \N_0$.  Let $a'=\theta_i^-(a)$ and
  $d'=\theta_i^-(d)$.  By Property P\ref{treitems-due}, $a$ and $a'$
  are compatible in $\N_i$, therefore $a'\at \Box_{\S} F\in \N_i$.
  Moreover, since $a'$ occurs in $\N_0$, by Property
  P\ref{treitems-uno}, $a'\at \Box_{\S}F\in \N_0$.  Since also $a'
\Rightarrow_{\R}d'\in \N_0$ and $\N_0$ is pseudo-saturated, $d'\at
  \Box_{\R}F\in \N_0$, so that also $\theta_i(d')\at \Box_{\R}F = d\at
  \Box_{\R}F\in \N_i$.
\qedhere
\end{enumerate}
\end{description}

\end{enumerate}
\end{proof}

The construction of a model of $\N_{\cal B}^\infty$ is here
substantially different from the corresponding one in \cite{jar2012},
and is inspired by the corresponding construction in \cite{HoSa99}.
In order to simplify the presentation, an intermediate result is
stated and proved next, based on the  following definition.

\begin{defi}\label{nuovadefi}
Let $\ramo$ be a complete and open branch.
For every relation symbol $\r$ occurring in \ramo, the following
notions are defined (in the notations used in the
first two items the branch $\ramo$ is left implicit):
\begin{enumerate}
\item $\r_{\subseteq} =\{\langle a,b\rangle\mid
a\Rightarrow_{\S} b\in\N_{\cal B}^\infty\mbox{ for some }\S\subrole
\r\in\ramo\}$, and
$\r_\subseteq^-=\{\langle a,b\rangle\mid \langle  b,a\rangle\in
  \r_\subseteq\}$. 

Moreover,
$\R_\subseteq$ is an abbreviation for $\r_\subseteq$ if $\R$ is a forward
  relation, otherwise it stands for $\r_\subseteq^-$.

\item $(\R_\subseteq)^+$ is the transitive closure of $\R_\subseteq$.


\item  $\rho_{\cal B}$ is the function on relation symbols 
defined as follows:
\[\rho_{\cal B}(\r) =
\left\{
\displaystyle{
\begin{array}{l}
(\r_\subseteq)^+ \mbox{~~~ if }\trans(\r)\in \ramo\\ 
\r_\subseteq\cup 
\{(\S_\subseteq)^+\mid 
\S\subrole \r\in{\cal B}
                      \mbox{ and }\trans(\S)\in\ramo\}
\\\mbox{ ~~~~~~~otherwise}
\end{array}}
\right.\]
$\rho_{\cal B}(\r^-)$ stands for $\{\langle a,b\rangle\mid \langle
  b,a\rangle\in \rho_{\cal B}(\r)\}$.
\end{enumerate}
\end{defi}

Below, the notation $\inv(\R)$ is used to denote
$\r^-$ if $\R=\r$ is a forward relation; 
otherwise, if $\R=\r^-$, then $\inv(\R)=\r$.

\begin{newlemma}\label{nuovolemma}
If $\ramo$ is a complete and open branch, then:
\begin{enumerate}
\item 
\label{uno-newlemma}
for every $\r\in\REL$ such that $\trans(\r)\in\ramo$, $\rho_{\cal B}(\r)$
  is a transitive relation;

\item \label{due} for every $\S\subrole \r\in\ramo$,
  $\S_\subseteq\subseteq \r_\subseteq$;

\item \label{nuovo1}
for every relation symbol $\r$ and nominals $a,b$,
if $\trans(\r)\in\ramo$, then   $\langle
  a,b\rangle \in\rho_{\cal B}(\r)$ if and only if there are nominals
  $c_0=a,c_1,\dots,c_n,c_{n+1}=b$ and relations $\P_0,\dots,\P_n$,
 for $n\geq 0$, such that $\P_i\subrole \r\in\ramo$ and
$c_i\Rightarrow_{\P_i} c_{i+1}\in\N_{\cal B}^\infty$, for all $i=0\dots
  n$.

\item 
\label{quattro-newlemma}
for every $\S\subrole \r\in\ramo$, $\rho_{\cal B}(\S)\subseteq
  \rho_{\cal B}(\r)$.


\item \label{sette} 
For every relation
 $\R$ and nominals $a,b$, 
  if $\langle a,b\rangle \in\rho_{\cal B}(\R)$ then one of the
  following cases holds:
\begin{itemize}
\item $a\Rightarrow_{\S} b\in\N_{\cal B}^\infty$ for some $\S\subrole
\R\in\ramo$;

\item there exist relations $\P, \P_0,\dots,\P_n$ and nominals
  $c_0=a,\dots,c_{n+1}=b$ ($n\geq 0$), such that
  $\trans(\P),~\P\subrole \R\in \ramo$ and, for all $i=0\dots n$,
  $\P_i\subrole \P\in\ramo$ and $c_i\Rightarrow_{\P_i}
  c_{i+1}\in\N_{\cal B}^\infty$.

\end{itemize}
\end{enumerate}
\end{newlemma}

\begin{proof}
The first item follows directly from the definition of $\rho_{\cal
  B}$.

\begin{enumerate}
\setcounter{enumi}{1}

\item Let us assume that $\S\subrole \r\in\ramo$ and $\langle
  a,b\rangle\in \S_\subseteq$. Then $a\Rightarrow_{\P} b\in\N_{\cal
    B}^\infty$ for some $\P\subrole \S\in\ramo$.  Since $\P\subrole
  \S$ and $\S\subrole \r$ are both in $\ramo$ and $\ramo$ is complete,
  also $\P\subrole \r$ is in $\ramo$. Therefore $\langle a,b\rangle\in
  \r_\subseteq$.

\item 
Let us assume that  that there are nominals
$c_0=a,c_1,\dots,c_n,c_{n+1}=b$ and relations
$\P_0,\dots,\P_n$ such that $\P_i\subrole \r\in\ramo$ and
 $c_i\Rightarrow_{\P_i}c_{i+1}\in\N_{\cal B}^\infty$,
 for all $i=0\dots n$. Then, $\langle c_i,c_{i+1}\rangle\in
 \r_\subseteq$, by definition. If moreover
$\trans(\r)\in\ramo$, 
then $\rho_{\cal B}(\r)=(\r_\subseteq)^+$, therefore $\langle
a,b\rangle\in\rho_{\cal B}(\r)$.

For the other direction, let us assume that $\trans(\r)\in\ramo$ and
 $\langle a,b\rangle \in \rho_{\cal B}(\r)=(\r_\subseteq)^+$.  
Then there are  nominals
$c_0=a,c_1,\dots,c_n,c_{n+1}=b$, for $n\geq 0$, such that 
$\langle c_i,c_{i+1}\rangle \in \r_\subseteq$ for all $i=0\dots n$.
For each such $i$,  $c_i\Rightarrow_{\P_i} c_{i+1}\in \N_{\cal
  B}^\infty$ for some $\P_i\subrole \r$.

\item 
Let us assume that $\S\subrole \r\in\ramo$. 
We distinguish the following cases:
\begin{enumerate}
\item Both $\trans(\S)$ and $\trans(\r)$ are in $\ramo$. 
Then $\rho_{\cal B}(\S)\subseteq \rho_{\cal B}(\r)$ follows from item
\ref{due} and the definition of $\rho_{\cal B}$.

\item If $\trans(\r)\not\in\ramo$ and $\trans(\S)\in\ramo$, then
  $\rho_{\cal B}(\S)\subseteq 
  \rho_{\cal B}(\r)$ follows directly from the definition of
  $\rho_{\cal B}$.

\item Let us finally consider the case where
 $\trans(\S)\not\in\ramo$ and $\trans(\r)\in\ramo$, and let us assume
  that $\langle a,b\rangle\in\rho_{\cal B}(\S)$.  

  If $\langle   a,b\rangle\in \S_\subseteq$, then $\langle
  a,b\rangle\in \r_\subseteq\subseteq \rho_{\cal B}(\r)$ 
by item \ref{due}.

  If $\langle   a,b\rangle\not \in \S_\subseteq$, then 
  $\langle   a,b\rangle\in\rho_{\cal B}(\P)$ for some $\P$ such that 
  $\P\subrole \S,~\trans(\p)\in\ramo$. 

By item \ref{nuovo1}, there are nominals $c_0=a,\dots, c_n,c_{n+1}=b$
and relations $\P_0,\dots,\P_n$ such that $\P_i\subrole \P\in\ramo$
and
 $\N_{\cal B}^\infty$ contains $c_i\Rightarrow_{\P_i} c_{i+1}$ for
all $i=0\dots n$.  Since $\P_i\subrole \P$,
 $\P\subrole \S$ and $\S\subrole \r$ are all in
$\ramo$, the branch also contains $\P_i\subrole \r$ for all  $i=0\dots n$. 
Therefore, $\langle c_i,c_{i+1}\rangle\in \r_\subseteq$ for all 
$i=0\dots n$ and $\langle a,b\rangle\in (\r_\subseteq)^+=\rho_{\cal
  B}(\r)$.

\end{enumerate}

\item 
The cases of forward and backward relations are treated separately.

\begin{enumerate}
\item
\label{Rpositive}
 $\R=\r$ is a forward relation.  If $\trans(\r)\in\ramo$, then the
  second case holds, following from item \ref{nuovo1}, taking $\P=\P_i=\r$
  for all $i=0,\dots,n$ (since $\ramo$ is complete, it contains
  $\r\subrole \r$). 

Let us assume that $\trans(\r)\not\in\ramo$ and $\langle a,b\rangle
\in\rho_{\cal B}(\r)$.  If $\langle a,b\rangle \in \r_\subseteq$ then
the first case holds because $\r_\subseteq\subseteq \rho_{\cal B}(\r)$.
If $\langle a,b\rangle \not\in \r_\subseteq$, then $\langle a,b\rangle
\in\rho_{\cal B}(\P)$ for some $\P$ such that $\P\subrole \r\in\ramo$
and $\trans(\P)\in\ramo$.  Therefore the second case holds, following
from item \ref{nuovo1}.

\item If $\R=\r^-$ is a backward relation, 
then $\langle a,b\rangle\in\rho_{\cal B}(\R)$ if
  and only if $\langle b,a\rangle\in\rho_{\cal B}(\r)$.  From 
case \ref{Rpositive} above,
 it  follows that one of the following cases holds:

\begin{itemize}

\item $b\Rightarrow_{\S}a\in\N^\infty_{\cal B}$ for some $\S\subrole
  \r\in \ramo$. 
Since $\S\subrole \r=\inv(\S)\subrole \r^-$ and
$b\Rightarrow_{\S}a=a\Rightarrow_{\inv({\S})}b$, the first case holds.

\item there exist relations $\P,\P_0,\dots,\P_n$ and nominals
  $c_0=a,\dots,c_{n+1}=b$, such that $\P\subrole
  \r\comma~\trans(\P)\in\ramo$ and, for all $i=0\dots n$, $\P_i\subrole \P
  \in\ramo$ and $c_{i+1}\Rightarrow_{\P_i} c_{i}\in\N_{\cal
    B}^\infty$.  Then the second case holds because $\P\subrole
  \r=\inv(\P)\subrole \r^-$,  $\P_i\subrole
  \P=\inv(\P_i)\subrole \inv(\P)$ and $c_{i+1}\Rightarrow_{\P_i} c_{i} =
  c_i\Rightarrow_{\inv(\P_i)} c_{i+1}$.\qedhere
\end{itemize}
\end{enumerate}
\end{enumerate}
\end{proof}

\setcounter{lemma}{12}
\begin{lemma}
\label{costruzione-modello}

If $\ramo$ is a complete and open branch, then
the possibly infinite set $\N_{\cal B}^\infty$ has a model.

\end{lemma}

\begin{proof}

Let ${\cal M}=\langle W,\rho, N, I\rangle$ be defined as follows: 
\begin{itemize}
\item
$W$ is the set of all the nominals occurring in $\N_{\cal B}^\infty$;
\item 
$\rho=\rho_{\cal B}$.
\item
$N(a)=a$ for every nominal $a$;
\item for any $p \in
\PROP$, $p \in I(a)$ if and only if $a\at p$ is the label of some node
in $ \N_{\cal B}^\infty$.
\end{itemize}

The fact that $\M$ is a model of the set of assertions in
$\ramo$ follows from Lemma \ref{nuovolemma} (items \ref{uno-newlemma}
and \ref{quattro-newlemma}).  

Next we prove that, for every $a\at F\in \N_{\cal B}^\infty$,
$\M_a\models F$.  The proof is by induction on $F$. All cases are
straightforward consequences of the definition of $\M$,
Lemma \ref{saturation_property} and the fact
that every blockable node has its witness(es) in 
$\N_{\cal B}^\infty$, except
for the case where $F=\Box_{\R} G$, whose treatment is
shown below.

The notation $F_1,\dots,F_n\allora^{k({\cal R})} F$ will be used to
mean 
that from the fact that $F_1,\dots,F_n\in \N_{\cal B}^\infty$ it can
be inferred that $F\in \N_{\cal B}^\infty$, because, by Lemma 
\ref{saturation_property}, $\N_{\cal B}^\infty$ satisfies item $k$
 of Definition \ref{pseudosat}, corresponding to the expansion rule
 $\cal R$. 
\begin{description}
\item [$(\Box_{\R})$] Let us assume that $a\at\Box_{\R} G\in \N_{\cal
  B}^\infty$.  It must be shown that $\M_b\models G$ for every
  $b$ such that $\langle a,b\rangle\in \rho_{\cal B}(\R)$. 

By item \ref{sette} of Lemma \ref{nuovolemma}, if $\langle
a,b\rangle\in \rho_{\cal B}(\R)$, then one of the cases that follow holds. For
each of them we show that $b\at G\in\N^\infty$. Then 
 $\M_b\models G$ follows from 
the
induction hypothesis.

\begin{enumerate}
\item $a\Rightarrow_{\S} b\in\N_{\cal B}^\infty$ for some $\S\subrole
\R\in\ramo$.
Then $b\at G\in\N^\infty$ because:
\[
\begin{array}{rll}
a\Rightarrow_{\S} b,~\S\subrole \R&\allora^{\ref{link}(\link)}~&
a\Rightarrow_{\R} b\\
a\at\Box_\R G,~a\Rightarrow_{\R} b&\allora^{\ref{multibox}(\Box)}~&
b\at G
\end{array}
\]

\item 
there exist relations $\P, \P_0,\dots,\P_n$ and nominals
  $c_0=a,\dots,c_{n+1}=b$ ($n\geq 0$), such that
  $\trans(\P),~\P\subrole \R\in \ramo$ and, for all $i=0\dots n$,
  $\P_i\subrole \P\in\ramo$ and $c_i\Rightarrow_{\P_i}
  c_{i+1}\in\N_{\cal B}^\infty$.
Then:
\[\begin{array}{rll}
a\Rightarrow_{\P_0}c_1,~\P_0\subrole\P &\allora^{\ref{link}(\link)}~&
a\Rightarrow_{\P}c_1
\\
a\at\Box_\R G,~a\Rightarrow_{\P}c_1,~
          \trans(\P),~\P\subrole \R & 
\allora^{\ref{trans}(\trans)}~
 & c_1\at\Box_{\P} G\\
c_1\Rightarrow_{\P_1}c_2,~\P_1\subrole\P &\allora^{\ref{link}(\link)}~&
c_1\Rightarrow_{\P}c_2\\
c_1\at\Box_\P G,
~c_1\Rightarrow_{\P}c_2,~
          \trans(\P),~\P\subrole \R & 
\allora^{\ref{trans}(\trans)}~
 & c_2\at\Box_{\P} G\\
\dots&\allora^{\ref{trans}(\trans)}~&
    c_n\at\Box_{\P} G\\
c_n\Rightarrow_{\P_n}b,~\P_n\subrole\P &\allora^{\ref{link}(\link)}~&
c_n\Rightarrow_{\P}b\\
c_n\at\Box_{\P} G, ~
c_n\Rightarrow_{\P}b
&\allora^{\ref{multibox}(\Box)}~ & b\at G
\end{array}
\]
\end{enumerate}

\end{description}
\end{proof}

Completeness can finally be proved using Lemma
\ref{costruzione-modello} like in \cite{jar2012}.

\begin{theorem}[Completeness]
Let $F$ be a formula and $\calA$ a set of assertions.  If
$\{F\}\cup\calA$ is in $\framcalcolo$ and is unsatisfiable, then any
complete tableau for $\{F\}\cup\calA$ is closed.
\end{theorem}